\documentclass[prd,aps,nofootinbib,10pt]{revtex4}
\usepackage{amsmath,amssymb,graphicx,epsfig}

\usepackage{graphics}
\usepackage{rotating}
\usepackage{epsfig}
\usepackage{color}

\begin{document}

\title{Chiral dynamics,  S-wave contributions  and angular analysis in  $D\to \pi\pi \ell\bar\nu$ }
\author{
Yu-Ji Shi, Wei Wang~\footnote{Email:wei.wang@sjtu.edu.cn} and Shuai Zhao~\footnote{Email:shuai.zhao@sjtu.edu.cn} }
\affiliation{
INPAC, Shanghai Key Laboratory for Particle Physics and Cosmology, Department of Physics and Astronomy, Shanghai Jiao-Tong University, Shanghai, 200240,   China}

\begin{abstract}
We present a theoretical analysis of  the $D^-\to \pi^+\pi^-
\ell\bar\nu$ and  $\bar D^0\to \pi^+\pi^0 \ell\bar\nu$ decays. We
construct a general  angular distribution which can include
arbitrary partial waves of $\pi\pi$. Retaining the S-wave and P-wave
contributions we study the branching ratios, forward--backward
asymmetries and a few other observables. The P-wave contribution is
dominated by $\rho^0$ resonance, and the S-wave contribution is
analyzed using the unitarized chiral perturbation theory.  The
obtained branching fraction for $D\to \rho\ell\nu$,  at the order
$10^{-3}$, is consistent with the available experimental data. The
S-wave contribution has a branching ratio  at the order of
$10^{-4}$, and this prediction can be tested  by experiments like
BESIII and LHCb. Future measurements can also be used to examine the
$\pi$--$\pi$ scattering phase shift.
\end{abstract}

\maketitle

\section{Introduction}

{ The Cabbibo--Kobayashi--Maskawa (CKM) matrix elements are key
parameters in the Standard Model (SM). They are essential to
understand CP violation within the SM and search for new physics
(NP). Among these matrix elements,
 $|V_{cd}|$ can
be determined from either exclusive or inclusive weak $D$ decays,
which are governed by $c\to d$ transition, for example, $c\to d\ell
\nu$ transitions. However, for a general $D$ decay process it is
difficult to extract CKM matrix elements, because strong and weak
interactions may be entangled.

 The semi-leptonic $D$ decays are ideal
channels to determine $|V_{cd}|$, not only because the weak and
strong dynamics can be separated in these process, but also the
clean experimental signals. Moreover, one can study the dynamics in
the heavy-to-light transition from semi-leptonic $D$ decays. For
leptons do not participate in the strong interaction, all the strong
dynamics is included in the form factors; thus it provides a good
platform to measure the form factors. The $D\to \rho$ form factors
have been measured from $D^0\to \rho^- e^+ \nu_e$ and $D^+\to \rho^0
e^+ \nu_e$ at the CLEO-c experiment for both charged and neutral
channels \cite{CLEO:2011ab}. Because of the large width of the
$\rho$ meson, $D\to\rho\ell\bar\nu_{\ell}$ is in fact a quasi-four
body process $D\to\pi\pi\ell\bar\nu_{\ell}$.  The  $\rho$ can be
reconstructed from the P-wave $\pi\pi$ mode.  However,  other
$\pi\pi$ resonant or non-resonant states may interfere with the
P-wave $\pi\pi$ pair, and thus  it is necessary to analyze the
S-wave contribution to $D\to \pi\pi\ell\bar\nu_{\ell}$.

In addition, the internal structure of light mesons is an important
issue in hadron physics. It is difficult to study light mesons by
QCD perturbation theory due to the large strong coupling in the low
energy region. On the other hand, because of the large mass scale,
one can establish factorization for many heavy meson decay
processes, thus heavy mesons like $B$ and $D$ can be used to probe
the internal structure of light
mesons~\cite{Wang:2009azc,Achasov:2012kk}.  As mentioned above,
$D\to \pi \pi \ell \bar\nu_{\ell}$ can receive contributions from
various partial waves of $\pi\pi$. $\rho$(770) dominant for $D$ to
P-wave $\pi\pi$ decay, at the same time, $D$ meson can decay into
S-wave $\pi\pi$ through $f_0(980)$. The structure of $f_0$(980) is
not fully understood  yet. Analysis of $D\to\pi\pi\ell
\bar\nu_{\ell}$ may shed more light on understanding the nature of
$f_0$(980). The BESIII collaboration has collected 2.93
$\mathrm{fb}^{-1}$ data in $e^+e^-$ collisions at the energy around
$3.773$ GeV \cite{Ablikim:2015orh}, which can be used to study the
semi-leptonic $D$ decays. Thus it presently is mandatory to make
reliable theoretical predictions. Some analyses of multi-body heavy
meson decays can be found in
Refs.~\cite{Lu:2011jm,Meissner:2013pba,Meissner:2013hya,Wang:2015uea,Wang:2015paa,Shi:2015kha,Xie:2014tma,Sekihara:2015iha,Oset:2016lyh,Wang:2016rlo,Kang:2013jaa,Faller:2013dwa,Niecknig:2015ija,Daub:2015xja,Albaladejo:2016mad},
where the final state interactions between the light pseudoscalar
mesons are taken into account.

In this paper we present a theoretical analysis of
$D^-\to\pi^+\pi^-\ell\bar \nu_{\ell}$ and $D^0\to \pi^+\pi^0\ell
\bar\nu_{\ell}$ decays.  In Sec.~II, we will present the results of
$D\to f_0$ (980) and $D\to \rho$ form factors. We also calculate $D$
to S-wave $\pi\pi$ pair form factors in non-resonance region, the
$\pi\pi$ form factor will be calculated by using unitarized chiral
perturbation theory. Based on these results, we present a full
analysis on the angular distribution of $D\to \pi\pi\ell \bar
\nu_{\ell}$. We explore various distribution observables, including
the differential decay width, the S-wave fraction, forward--backward
asymmetry, and so on. These results will be collected in Sec.~III.
The conclusion of this paper will be given in Sec.~IV. The details
of the coefficients in angular distributions are relegated to the
appendix.

}

\section{Heavy-to-light transition form factors}
\label{sec:formfactor}

\begin{figure}\begin{center}
\includegraphics[scale=0.6]{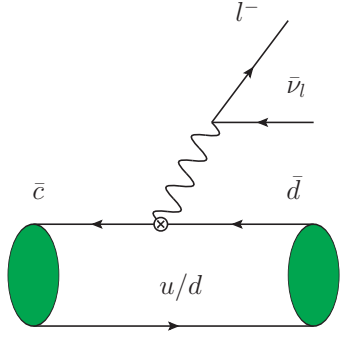}
\caption{Feynman diagram for the $D\to \pi\pi \ell^-\bar\nu_{\ell}$
decays.   The lepton could be an electron or a muon, $\ell=e,\mu$.
Depending on the $D$ meson, the spectator could be a $u$ or a $d$
quark, corresponding to $D^0\to \pi^+\pi^0 \ell^-\bar \nu_{\ell}$
and $D^-\to \pi^+\pi^- \ell^-\bar \nu_{\ell}$  } \label{fig:feynman}
\end{center}
\end{figure}

Feynman diagram for the $D\to \pi\pi \ell^-\bar\nu_{\ell}$ decay is
shown in Fig.~\ref{fig:feynman}. The lepton can  be an  electron or
a muon, $\ell=e,\mu$. The spectator quark could be the $u$ or $d$
quark, corresponding to $D^0\to \pi^+\pi^0 \ell^-\bar \nu_{\ell}$
and $D^-\to \pi^+\pi^- \ell^-\bar \nu_{\ell}$. Integrating out the
virtual $W$-boson,  we  obtain  the effective Hamiltonian describing
the $c\to  d$ transition
\begin{eqnarray}
 \mathcal{H}_{\mathrm{eff}} = \frac{G_{\mathrm{F}}}{\sqrt 2} V_{cd} [\bar d \gamma_\mu(1-\gamma_5) c][ \bar \nu \gamma^\mu(1-\gamma_5) \ell] +h.c.,
\end{eqnarray}
where $G_\mathrm{F}$ is the Fermi constant and $V_{cd}$ is the CKM
matrix element. The leptonic part is calculable using the
perturbation theory, while the hadronic effects are encoded into the
transition form factors.

\subsection{$D \to \rho$ form factors}
For the P-wave $\pi\pi$ state, the dominant contribution is from
the $\rho(770)$ resonance. The  $D\to \rho$ form factors are
parametrized  by~\cite{Wirbel:1985ji}
 \begin{eqnarray}
  \langle \rho(p_2,\epsilon)|\bar d\gamma^{\mu}c|D(p_D)\rangle
   &=&-\frac{2V(q^2)}{m_{D}+m_{\rho}}\epsilon^{\mu\nu\rho\sigma}
   \epsilon^*_{\nu}  p_{D\rho}p_{2\sigma}, \nonumber\\
  \langle  \rho(p_2,\epsilon)|\bar d\gamma^{\mu}\gamma_5 c|
  D(p_D)\rangle
   &=&2im_{{\rho}} A_0(q^2)\frac{\epsilon^*\cdot  q }{ q^2}q^{\mu}
    +i(m_{D}+m_{{\rho}})A_1(q^2)\left[ \epsilon^{*\mu }
    -\frac{\epsilon^*\cdot  q }{q^2}q^{\mu} \right] \nonumber\\
    &&-iA_2(q^2)\frac{\epsilon^* \cdot  q }{  m_{D}+m_{{\rho}} }
     \left[ P^{\mu}-\frac{m_{D}^2-m_{{\rho}}^2}{q^2}q^{\mu} \right],
     \label{eq:BtoTformfactors-definition}
 \end{eqnarray}
{ with $q=p_D-p_2$, and $P=p_D+p_2$.}  The $V(q^2)$, and
$A_{i}(q^2) (i=0,1,2)$ are nonperturbative form factors.

These form factors have been computed in many different approaches
{  \cite{Scora:1995ty,Fajfer:2005ug,Verma:2011yw,Cheng:2003sm,Wu:2006rd}},
and here we quote the results from the light-front quark model
(LFQM)~\cite{Cheng:2003sm,Verma:2011yw} and light-cone sum rules (LCSR)
~\cite{Wu:2006rd}. To access the momentum distribution in the full
kinematics region,  the following parametrization has been used:
\begin{equation}
F_{i}(q^{2})=\frac{F_{i}(0)}{1-a_{i}\frac{q^{2}}{m_{D}^{2}}+b_{i}\left(\frac{q^{2}}{m_{D}^{2}}\right)^{2}}.
\label{eq:formfactorParametrization}
\end{equation}
Their results are  collected in Tab.~\ref{Tab:DtoRhoformFactor}. We
note that a different parametrization is adopted in
Ref.~\cite{Wu:2006rd}, where $A_3$ appears instead of $A_0$. The
relation between $A_0$ and $A_3$ is given by
\begin{eqnarray}
A_{0}(q^{2})=\frac{1}{2m_{\rho}(m_{D}+m_{\rho})}\left[A_{1}(q^{2})
(m_{D}+m_{\rho})^{2}+A_{2}(q^{2})(m_{\rho}^{2}-m_{D}^{2})-A_{3}(q^{2})q^{2}\right].
\end{eqnarray}


\begin{table}
  \caption{$D \to \rho$ form factors derived by
LFQM (left) \cite{Verma:2011yw,Cheng:2003sm} and LCSR (right)
\cite{Wu:2006rd}, respectively} \label{Tab:DtoRhoformFactor}
\begin{center}
\begin{tabular}{|c|c|c|c||c|c|c|c|}
\hline $\mathrm{LFQM}$ & $F(0)$ & $a$ & $b$ & $\mathrm{LCSR}$ &
$F(0)$ & $a$ & $b$\tabularnewline \hline

$V^{D\rightarrow\rho}$ & $0.88\pm 0.03$ & $1.23\pm 0.01$ & $0.40\pm
0.04$ & $V^{D\rightarrow\rho}$ & $0.801\pm 0.044$ & $0.78\pm 0.24$ &
$2.61\pm 0.29$\tabularnewline

$A_{0}^{D\rightarrow\rho}$ & $0.69\pm 0.02$ & $1.08\pm 0.02$ &
$0.45\pm 0.04$ & $A_{3}^{D\rightarrow\rho}$ & $-0.719\pm 0.066$ &
$1.05\pm 0.15$ & $1.77\pm 0.20$\tabularnewline

$A_{1}^{D\rightarrow\rho}$ & $0.60\pm 0.01$ & $0.46\pm 0.03$ &
$0.01\pm 0.00$ & $A_{1}^{D\rightarrow\rho}$ & $0.599\pm 0.035$ &
$0.44\pm 0.10$ & $0.58\pm 0.23$\tabularnewline

$A_{2}^{D\rightarrow\rho}$ & $0.47\pm 0.00$ & $0.89\pm 0.02$ &
$0.23\pm 0.03$ & $A_{2}^{D\rightarrow\rho}$ & $0.372\pm 0.031$ &
$1.64\pm 0.16$ & $0.56\pm 0.28$\tabularnewline \hline
\end{tabular}
\end{center}
\end{table}

\subsection{Scalar $\pi\pi$ form factor and $D$ to S-wave $\pi\pi$}
\label{sec:pipiff}

We first give the $D\to f_0(980)$ form factor parametrized as
\begin{eqnarray}
    \langle f_0(p_2) |\bar d  \gamma_\mu\gamma_5  c|
    {D^-}(p_{D})\rangle
    &=&-i\left\{    F_+^{D\to f_0} (q^2)\left
    [P_\mu - \frac{m_{D}^2-m_{f_0}^2}{q^2}q_\mu \right ]
    +F_0^{D\to f_0} (q^2)\frac{m_{D}^2-m_{f_0}^2}{q^2}q_\mu
    \right\},
  \end{eqnarray}
where  {  $F^{D\to f_0}_+$ and $F^{D\to f_0}_0$ are $D\to f_0$ form
factors.} We will use LCSR to compute the   $D \to f_0(980)$
transition  form factors with some inputs, and we refer the reader
to Ref.~\cite{Colangelo:2010bg} for a detailed derivation in LCSR.
The meson masses are fixed to the PDG values {  $m_D=1.870$\ GeV and
$m_{f_0}=0.99$ \ GeV~\cite{Olive:2016xmw}}. For quark masses we use
{ $m_c=1.27$\ GeV~\cite{Olive:2016xmw} and $m_d=5$\ MeV}. As for
decay constants, we use $f_D=0.21$\ GeV~\cite{Olive:2016xmw} and
$f_{f_0}=0.18$\ GeV~\cite{DeFazio:2001uc}. The threshold $s_0$ is
fixed at $s_0=4.1$\ GeV$^2$, which should correspond to the squared
mass of the first radial excitation of $D$. { The parameters
$F_i(0)$, $a_i$ and $b_i$ are fitted in the region
$-0.5~\mathrm{GeV^2}<q^2<0.5~\mathrm{GeV^2}$, and the Borel
parameter $M^2$ is taken to be $(6\pm1)~\mathrm{GeV}^{-2}$}. With
these parametrizations, we give the numerical results in
Tab.~\ref{Tab:Dtof0formFactor}.


\begin{table}
\caption{Fitted parameters of the $D \to f_0$ form factors derived
by LCSR, which are fitted using
Eq.~\eqref{eq:formfactorParametrization}}
\label{Tab:Dtof0formFactor}
\begin{center}
\begin{tabular}{|c|c|c|c|}
\hline $D\to f_{0}$ & $F(0)$ & $a$ & $b$\tabularnewline \hline
$F_{1}$ & $0.321\pm 0.010$ & $0.990\pm 0.032$ & $0.543\pm 0.023
$\tabularnewline $F_{0}$ & $0.321\pm 0.010$ & $0.344\pm 0.019$ &
$-0.735\pm 0.001$\tabularnewline \hline
\end{tabular}
\end{center}
\end{table}


In the region where the two pseudo-scalar mesons strongly interacts, the resonance approximation fails and thus has to be abandoned. One of the such examples is the S-wave partial wave under $1$ GeV, for which we can use the form factors as defined  in Ref.~\cite{Doring:2013wka}:
\begin{eqnarray}
 \langle (\pi\pi)_S(p_{\pi\pi})|\bar u \gamma_\mu\gamma_5 c| D (p_D)
 \rangle  &=& -i  \frac{1}{m_{\pi\pi}} \bigg\{ \bigg[P_{\mu}
 -\frac{m_{D}^2-m_{\pi\pi}^2}{q^2} q_\mu \bigg] {\cal F}_{1}^{D\to \pi\pi}(m_{\pi\pi}^2, q^2) \nonumber\\
 &&
 +\frac{m_{D}^2-m_{\pi\pi}^2}{q^2} q_\mu  {\cal F}_{0}^{D\to \pi\pi}(m_{\pi\pi}^2, q^2)  \bigg\}.
 \label{eq:generalized_form_factors}
\end{eqnarray}

The Watson theorem implies that phases measured in $\pi\pi$ elastic
scattering and in a decay channel in which the $\pi\pi$ system has
no strong interaction with other hadrons  are equal modulo $\pi$
radians. In the process we consider here, the lepton pair $\ell
\bar\nu$ indeed decouple{s from} the $\pi\pi$ final state, and thus
the phases of $D$ to scalar $\pi\pi$  decay amplitudes  are equal to
$\pi\pi$ scattering with the same isospin. It is plausible that
\begin{eqnarray}
 \langle (\pi\pi)_S|\bar d \Gamma c|D\rangle
 \propto F_{\pi\pi}(m_{\pi\pi}^2),
\end{eqnarray}
where  the scalar form factor is defined as
\begin{eqnarray}
 \langle 0| \bar dd |\pi^+\pi^-\rangle = B_0
 F_{\pi\pi}(m_{\pi\pi}^2),
\label{defff}
\end{eqnarray}
{ where $B_0=(1.7\pm0.2)$ GeV \cite{Shi:2015kha} is the QCD condensate
parameter. }

An explicit calculation of these quantities  requires knowledge of
generalized light-cone distribution amplitudes
(LCDAs)~\cite{Diehl:2003ny}. The twist-3 one  has the same
asymptotic form with the LCDAs  for a scalar
resonance~\cite{Cheng:2005nb}.   Inspired by this similarity, we may
plausibly  introduce an intuitive matching between the $D\to f_0$
and $D\to (\pi\pi)_{S}$ form factors~\cite{Meissner:2013hya}:
\begin{eqnarray}
 {\cal F}_i^{D\to \pi\pi}(m_{\pi\pi}^2, q^2) \simeq  B_0\frac{1}{ f_{f_0}}  F_{\pi\pi}(m_{\pi\pi}^2) F_{i}^{D\to f_0}
 (q^2). \label{eq:matching}
\end{eqnarray}

It is necessary to stress at this stage that the Watson theorem does
not strictly guarantee that one may use Eq.~\eqref{eq:matching}.
Instead it indicates that, below the opening of inelastic channels
the strong phases in the $D\to\pi\pi$ form factor and $\pi\pi$
scattering are the same. First above the $4\pi$ or $K\bar K$
threshold, additional inelastic channels will also contribute. The
$K\bar K$ contribution can be incorporated in a coupled-channel
analysis.  As a process-dependent study, it has been demonstrated
that states with two additional pions may not give sizable
contributions to the physical observables~\cite{Bar:2012ce}.
Secondly, some polynomials with nontrivial  dependence on
$m_{\pi\pi} $ have been neglected in Eq.~\eqref{eq:matching}. In
principle, once the generalized LCDAs for the $(\pi\pi)_S$ system
are known,  the $D\to \pi\pi$ form factor can be straightforwardly
calculated in LCSR and thus this approximation in the matching
equation can be avoided. On the one side,   the space-like
generalized parton distributions for the pion have been calculated
at one-loop level  in the chiral perturbation theory
($\chi$PT)~\cite{Diehl:2005rn}. The analysis of time-like
generalized LCDAs in $\chi$PT and the  unitarized framework is in
progress.  On the other  side, the $\gamma\gamma^*\to \pi^+\pi^-$
reaction is helpful to extract the generalized LCDAs for the
$(\pi\pi)_S$ system~\cite{Diehl:1998dk,Diehl:2000uv}. The
experimental prospects at BEPC-II and BELLE-II in the near future
are very promising.

In the kinematic region where the $\pi$ is soft, the crossed channel
from $D+\pi\to \pi$ will contribute as well and this crossed channel
would modify Eq.~(\ref{eq:matching}) by an inhomogeneous part. For
the analogous decay of $K$ or $B$ mesons, it has been taken into
account either dynamically in terms of phase shifts (in the case of
the kaon decay)~\cite{Colangelo:2015kha} or approximately in terms
of a pole contribution (in the case of the $B$ meson
decay)~\cite{Kang:2013jaa}. However, if both pions move fast, the
$D$--$\pi$ invariant mass is far from the $D^*$ pole and this
contribution is negligible. In this case,  the transition amplitude
for the $D$ to 2-pion form factor can be calculated in light-cone
sum rules~\cite{Meissner:2013hya}. This will lead to the conjectured
formula in Eq.~\eqref{eq:matching}.

The scalar $\pi\pi$ form factor can be handled using the unitarized chiral perturbation theory. In the following, we will give a brief description of this approach.
In terms of  the isoscalar $S$-wave states
\begin{eqnarray}
&|\pi\pi\rangle_{\mathrm{I=0}}^{} & = \frac{1}{\sqrt{3}}
\left|\pi^+\pi^-\right\rangle + \frac{1}{\sqrt{6}}
\left|\pi^0\pi^0\right\rangle, \\
&|K\bar K\rangle_{\mathrm{I=0}}^{} & =
\frac{1}{\sqrt{2}}\left|K^+K^-\right\rangle +
\frac{1}{\sqrt{2}}\left|K^0\bar K^0\right\rangle,
\end{eqnarray}
the scalar form factors for the $\pi$ and $K$  mesons are defined as
\begin{eqnarray}
\sqrt{2}B_0\, F^{n/s}_1(s) &=& \langle 0|\bar nn /\bar ss|\pi\pi
\rangle_{\mathrm{I=0}}^{},
\label{FFdef} \\
\sqrt{2}B_0\,F^{n/s}_2(s) &=& \langle 0|\bar nn/\bar ss|K\bar K
\rangle_{\mathrm{I=0}}^{}, \nonumber
\end{eqnarray}
where $s=m_{\pi\pi}^2$. The $\bar nn = (\bar uu+\bar dd)/\sqrt2$
denotes the non-strange scalar current, and  the notation ($\pi$ =
1, $K$ = 2) has been introduced  for simplicity. With the above
notation, we have
\begin{eqnarray}
F_{\pi\pi}(m_{\pi\pi}^2)  = \sqrt{\frac{2}{3}} F_1^n(m_{\pi\pi}^2).
\end{eqnarray}
 Expressions  have already been derived
in $\chi$PT  up to next-to-leading
order~\cite{Gasser:1983yg,Gasser:1984gg,Gasser:1984ux,Meissner:2000bc}:
\begin{eqnarray}
&F_1^n(s)\:\: = & \left. \sqrt{\frac{3}{2}} \right[ 1 + \mu_\pi -
\frac{\mu_\eta}{3} + \frac{16 m_\pi^2}{f^2}\left(2L_8^r-L_5^r\right)
+ 8\left(2L_6^r-L_4^r\right)\frac{2m_K^2 + 3m_\pi^2}{f^2} +
\frac{8s}{f^2} L_4^r + \frac{4s}{f^2} L_5^r \nonumber \\ && + \left.
\left(\frac{2s - m_\pi^2}{2f^2}\right) J^r_{\pi\pi}(s) +
\frac{s}{4f^2} J^r_{KK}(s) + \frac{m_\pi^2}{18f^2} J^r_{\eta\eta}(s)
\right],   \\
&F_1^s(s)\:\: = & \frac{\sqrt{3}}{\:2} \left[ \frac{16
m_\pi^2}{f^2}\left(2L_6^r-L_4^r\right) + \frac{8s}{f^2} L_4^r +
\frac{s}{2f^2} J^r_{KK}(s) + \frac{2}{9}\frac{m_\pi^2}{f^2}
J^r_{\eta\eta}(s)
\right],\\
&F_2^n(s)\:\: = & \left. \frac{1}{\sqrt{2}} \right[ 1 +
\left.\frac{8 L_4^r}{f^2} \left(2s - m_\pi^2 - 6 m_K^2\right) +
\frac{4 L_5^r}{f^2} \left(s - 4 m_K^2\right) + \frac{16 L_6^r}{f^2}
\left(6 m_K^2 + m_\pi^2\right) + \frac{32 L_8^r}{f^2}\,m_K^2 +
\frac{2}{3} \mu_\eta \right. \nonumber \\ && + \left. \left(\frac{9s
- 8 m_K^2}{36f^2}\right) J^r_{\eta\eta}(s) + \frac{3s}{4f^2}
J^r_{KK}(s) + \frac{3s}{4f^2} J^r_{\pi\pi}(s)
\right], \\
&F_2^s(s)\:\: = & 1 + \frac{8 L_4^r}{f^2} \left(s - m_\pi^2 - 4
m_K^2\right) + \frac{4 L_5^r}{f^2} \left(s - 4 m_K^2\right) +
\frac{16 L_6^r}{f^2} \left(4 m_K^2 + m_\pi^2\right) + \frac{32
L_8^r}{f^2}\,m_K^2 + \frac{2}{3} \mu_\eta \nonumber \\ && +
\left(\frac{9s - 8 m_K^2}{18f^2}\right) J^r_{\eta\eta}(s) +
\frac{3s}{4f^2} J^r_{KK}(s).
\end{eqnarray}
Here the $L_i^r$ are the renormalized low-energy constants, and $f$
is the pion decay constant at tree level. The $\mu_i$ and $J_{ii}^r$
are defined as follows:
\begin{eqnarray}
\mu_i &=& \frac{m_i^2}{32\pi^2 f^2} \ln \frac{m_i^2}{\mu^2},\\
J_{ii}^r(s) \!&=&\! \frac{1}{16\pi^2}\left[ 1 -
\log\left(\frac{m_i^2}{\mu^2}\right) - \sigma_i(s)\log\left(
\frac{\sigma_i(s)+1}{\sigma_i(s)-1}\right)\right],
\end{eqnarray}
with  $ \sigma_i(s) = \sqrt{1- {4m_i^2}/{s}}$. It is interesting to
note that the next-to-next-to-leading order results can also be
found in Refs.~\cite{Bijnens:1998fm,Bijnens:2003xg}. Imposing  the
unitarity constraints, the scalar form factor can be expressed  in
terms of the algebraic  coupled-channel equation
\begin{eqnarray}
F(s) &=& [I+K(s)\,g(s)]^{-1} R(s) \label{G_eq} \\
&=& [I-K(s)\,g(s)]\:R(s) \:+\: \mathcal{O}(p^6), \nonumber
\end{eqnarray}
where $R(s)$ has no right-hand cut and   in the second line, the
equation has been expanded up to NLO  in the  chiral expansion.
$K(s)$ is the $S$-wave  projected  kernel of meson-meson scattering
amplitudes that can be derived from the  leading-order chiral
Lagrangian:
\begin{eqnarray}
&& K_{11} = \frac{2s - m_\pi^2}{2f^2}, \quad
   K_{12} = K_{21} = \frac{\sqrt{3}s}{4f^2},  \;\;\;
   K_{22} = \frac{3s}{4f^2} \ .
\nonumber
\end{eqnarray}
 The  loop integral can be  calculated either in the
cutoff-regularization scheme with $q_{\rm max}\sim 1~$GeV being the
cutoff (cf. Erratum of Ref.~\cite{Oller:1998hw} for an explicit
expression) or  in dimensional regularization with the
$\overline{\rm MS}$ subtraction scheme. In the latter scheme,  the
meson loop function $g_i(s)$ is  given by
\begin{eqnarray}
J_{ii}^r(s) \!&\equiv&\! \frac{1}{16\pi^2}\left[ 1 -
\log\left(\frac{m_i^2}{\mu^2}\right) - \sigma_i(s)\log\left(
\frac{\sigma_i(s)+1}{\sigma_i(s)-1}\right)\right] \nonumber \\
&=& -g_i(s). \label{m_loop}
\end{eqnarray}

\begin{figure}\begin{center}
\includegraphics[scale=0.7]{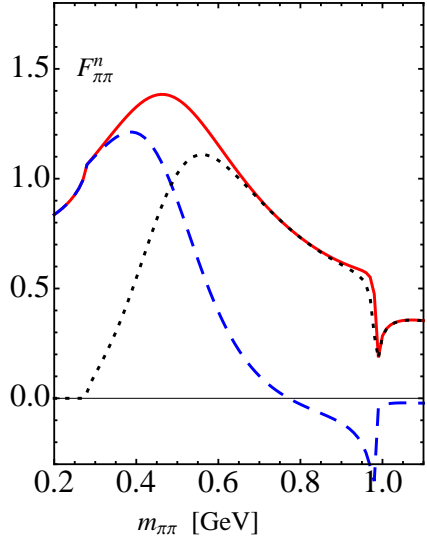}
\caption{The non-strange   $\pi\pi$      scalar form factor obtained
in the unitarized chiral perturbation theory. The modulus, real part
and imaginary part are shown in \textit{solid}, \textit{dashed} and
\textit{dotted curves} } \label{fig:pipi_ff}
\end{center}
\end{figure}
The expressions for the $R_i$ are obtained  by matching the
unitarization and chiral perturbation
theory~\cite{Oller:1997ti,Lahde:2006wr}:
\begin{eqnarray}
&R_1^n(s)\:\: = & \sqrt{\frac{3}{2}} \bigg\{ 1 + \mu_\pi -
\frac{\mu_\eta}{3} + \frac{16 m_\pi^2}{f^2}\left(2L_8^r-L_5^r\right)
+ 8\left(2L_6^r-L_4^r\right)\frac{2m_K^2 + 3m_\pi^2}{f^2}
+ \frac{8s}{f^2} L_4^r + \frac{4s}{f^2} L_5^r  \nonumber \\
&& -  \frac{m_\pi^2}{288\pi^2 f^2} \bigg[1 +
\log\left(\frac{m_\eta^2}{\mu^2}\right)\bigg]
\bigg\}, \\
&R_1^s(s)\:\: = & \frac{\sqrt{3}}{\:2} \left\{ \frac{16
m_\pi^2}{f^2}\left(2L_6^r-L_4^r\right) + \frac{8s}{f^2} L_4^r -
\frac{m_\pi^2}{72\pi^2 f^2} \left[1 +
\log\left(\frac{m_\eta^2}{\mu^2}\right)\right]
\right\},\\
&R_2^n(s)\:\: = & \frac{1}{\sqrt{2}} \left\{ 1 + \frac{8 L_4^r}{f^2}
\left(2s - 6m_K^2 - m_\pi^2\right) + \frac{4 L_5^r}{f^2} \left(s -
4m_K^2\right) + \frac{16 L_6^r}{f^2} \left(6m_K^2 + m_\pi^2\right)
+ \frac{32 L_8^r}{f^2} m_K^2 + \frac{2}{3} \mu_\eta \right.\nonumber \\
&& + \left.\frac{m_K^2}{72\pi^2 f^2} \left[1 +
\log\left(\frac{m_\eta^2}{\mu^2}\right)\right]
\right\}, \\
&R_2^s(s)\:\: = & 1 + \frac{8 L_4^r}{f^2} \left(s - 4m_K^2 -
m_\pi^2\right) + \frac{4 L_5^r}{f^2} \left(s - 4m_K^2\right) +
\frac{16 L_6^r}{f^2} \left(4m_K^2 + m_\pi^2\right)
+ \frac{32 L_8^r}{f^2} m_K^2 + \frac{2}{3} \mu_\eta \nonumber \\
&& + \frac{m_K^2}{36\pi^2 f^2} \left[1 +
\log\left(\frac{m_\eta^2}{\mu^2}\right)\right].
\end{eqnarray}

With the above formulas and the fitted results for the low-energy
constants $L_i^r$ in Ref.~\cite{Lahde:2006wr} (evolved  from
$M_\rho$ to the scale $\mu= 2q_{\rm max}/\sqrt{e}$), we show the
non-strange  $\pi\pi$   form factor  in Fig.~\ref{fig:pipi_ff}.  The
modulus, real part and imaginary part are shown as solid, dashed and
dotted curves. As the figure shows, the chiral unitary ansatz
predicts a form factor $F^n_{1}$ with a zero close to the $\bar KK$
threshold. This feature has been extensively discussed in
Ref.~\cite{Oller:2007xd}.

\section{Full angular distribution of $D\to \pi\pi  \ell \bar \nu$}
\label{sec:BstoKstar}

\begin{figure}\begin{center}
\includegraphics[scale=0.6]{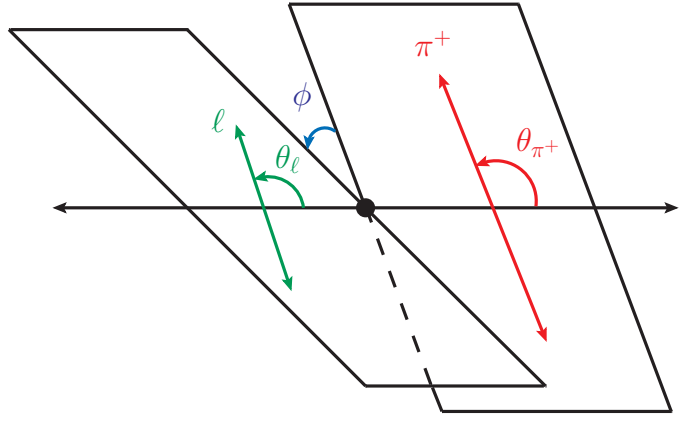}
\caption{Kinematics in the $D\to \pi\pi  \ell \bar \nu$. {The}
$\pi\pi $ moves along the $z$ axis in the $\overline D$ rest frame.
$\theta_\pi(\theta_{\ell})$ is defined in the $\pi\pi$ (lepton pair)
rest frame as the angle between $z$-axis and the flight direction of
$\pi^+$ ($\ell^-$), respectively. The azimuth angle $\phi$ is the
angle between the $\pi\pi$ decay and lepton pair planes }
\label{fig:kinematicsKpi}
\end{center}
\end{figure}

In this section, we will derive a full angular distribution of $D\to \pi\pi  \ell \bar \nu$. For the literature, one may consult Refs.~\cite{Cabibbo:1965zzb,Pais:1968zza}.
We set up the kinematics  for the $D^-\to \pi^+\pi^-  \ell \bar \nu$
as shown in Fig.~\ref{fig:kinematicsKpi}, which can also be used for   $\overline D^0\to \pi^+\pi^0  \ell \bar \nu$. The  $\pi\pi $ moves
along the $z$ axis in the $D^-$ rest frame.
$\theta_{\pi^+}(\theta_{\ell})$ is defined in the $\pi\pi$ (lepton
pair) rest frame as the angle between $z$-axis and the flight
direction of ${\pi}^+ $ ($\ell^-$), respectively. The azimuth angle
$\phi$ is the angle between the $\pi\pi$ decay and lepton pair
planes.

Decay amplitudes   for $D\to  \pi\pi \ell \bar \nu_{\ell}$ can
be divided into  several individual pieces and each of them can be
expressed in terms of the  Lorentz invariant helicity amplitudes.
The amplitude for the hadronic part can be obtained by   the
evaluation of the matrix element:
\begin{eqnarray}
A_{\lambda}  =   \sqrt{ N_{f_0/\rho}}   \frac{iG_{\mathrm{F}}}{\sqrt
2} V_{cd}^* \epsilon_{\mu}^*(h) \langle \pi\pi |\bar c
\gamma^\mu(1-\gamma_5) d |\overline  D\rangle,
\end{eqnarray}
where $\epsilon_\mu(h)$ is an auxiliary polarization vector for the
lepton pair system and $h= 0, \pm, t$,    $N_{f_0/\rho}=
\sqrt{\lambda} q^2\beta_l /(96 \pi^3m_{D}^3)$, $\beta_l=1-\hat
m_l^2$ and $\hat m_l= m_{l}/\sqrt{q^2}$. {  $|V_{cd}|$ is taken to
be 0.22 \cite{Olive:2016xmw}}. The functions $A_{i}$ can be
decomposed into different partial waves,
\begin{eqnarray}
 A_{0/t }(q^2, m_{\pi\pi}^2,\theta_{\pi^+})&=& \sum_{J=0,1,2...}  A^J_{0/t }(q^2, m_{\pi\pi}^2)Y_{J}^0(\theta_{\pi^+},0),\nonumber\\
 A_{||/\perp }(q^2, m_{\pi\pi}^2,\theta_{\pi^+})&=& \sum_{J= 1,2...}  A^J_{||/
 \perp }(q^2, m_{\pi\pi}^2)Y_{J}^{-1}(\theta_{\pi^+},0),\nonumber \\
 A^J_{0/t }(q^2, m_{\pi\pi}^2)&=&   \sqrt{ N_{f_0/{\rho}}} {\cal M}_D(f_0/{\rho}, 0/t )(q^2)
 L_{f_0/{\rho}}(m_{\pi\pi}^2) \equiv | A^J_{ 0/t }|
 e^{i\delta^J_{{ 0/t }}},\nonumber\\
 A^J_{ ||/\perp }(q^2, m_{\pi\pi}^2)&=&   \sqrt{ N_{f_0/{\rho}}} {\cal M}_D(f_0/{\rho},  ||/\perp)(q^2)
 L_{f_0/{\rho}}(m_{\pi\pi}^2)\equiv | A^J_{||/\perp}| e^{i\delta^J_{{ ||/
 \perp }}}.
\end{eqnarray}
 Here $J$ denotes the partial wave of the $\pi\pi$ system and   the script $t$ denotes the time-like component of a virtual vector/axial-vector meson decays into a lepton pair.  The  $L_{f_0/\rho}(m_{\pi\pi})$ is the   lineshape and  for
the P-wave $\rho$ we use the Breit--Wigner distribution:
\begin{eqnarray}
 L_{{\rho}}(m_{\pi\pi}^2)= \sqrt{ \frac{ m_{{\rho}} \Gamma_{{\rho}\to \pi\pi}  }{\pi} }
 \frac{1}{m_{\pi\pi}^2 -m_{{\rho}}^2+ i m_{{\rho}}\Gamma_{{\rho}}}.
\end{eqnarray}
Considering the  {momentum dependence of the $\rho$ decay},  we have
the running width as
\begin{eqnarray}
 \Gamma_{
 \rho} (m_{\pi\pi}^2) =  \Gamma_{{\rho}}^0 \left(\frac{ |\vec q\,|}{ |
 \vec q_0|}\right)^3   \frac{m_{{\rho}}}{m_{\pi\pi}}
 \frac{1+ (R|\vec q_0|)^2}{1+ (R|\vec q\,|)^2},\label{eq:KstarLineShape}
\end{eqnarray}
and the Blatt--Weisskopf parameter $R=(2.1\pm 0.5\pm 0.5)~ {\rm
GeV}^{-1}$~\cite{delAmoSanchez:2010fd}.

The spin-0 final state has only one polarization state and the
amplitudes are
\begin{eqnarray}
 i{\cal M}_D(f_0,0)&=& N_1 i\Bigg[   \frac{\sqrt {\lambda}}{\sqrt{ q^2}} F_1(q^2)  \Bigg],\;\;\;
 i{\cal M}_D(f_0,t)=N_1 i\Bigg[ \frac{m_{D}^2-m_{f_0}^2}{\sqrt {q^2}} F_0(q^2) \Bigg],
\end{eqnarray}
with  $N_1=  {iG_\mathrm{F}}V_{cd}^*/{\sqrt 2}$. For mesons with
spin $J\ge1$, the $\pi^+\pi^-$ system can be either longitudinally
or transversely polarized and thus we have the following form:
\begin{eqnarray}
 i{\cal M}_D(\rho,0)&=&-\frac{ \alpha_L^J N_1  i}{2m_{\rho}\sqrt {q^2}}\left[   (m_{D}^2-m_{\rho}^2-q^2)(m_{D}+m_{\rho})A_1
 -\frac{\lambda}{m_{D}+m_{\rho}}A_2\right], \nonumber\\
 i{\cal M}_D({\rho},\pm)
 &=& -\beta_T^J N_1  i \left[  (m_{D}+m_{\rho})A_1\pm \frac{\sqrt \lambda}{m_{D}+m_{\rho}}V \right],\\
 i{\cal M}_D({\rho},  t)&=&-\alpha_L^J i N_{1}  \frac{\sqrt \lambda}{\sqrt {q^2}}A_0.
\end{eqnarray}
The $\alpha_L^J$ and $\beta_T^J$ are products of the Clebsch--Gordan
coefficients
\begin{eqnarray}
 \alpha_L^J &=& C^{J,0}_{1,0;J-1,0} C^{J-1,0}_{1,0; J-2,0} \cdots C^{2,0}_{1,0;1,0},\;\;\;
 \beta_T^J = C^{J,1}_{1,1;J-1,0} C^{J-1,0}_{1,0; J-2,0} \cdots C^{2,0}_{1,0;1,0}.
\end{eqnarray}

For the sake of convenience, we define
\begin{eqnarray}
 i{\cal M}_{D}(\rho, \perp/||)&=&\frac{1}{\sqrt 2}[i{\cal M}_D(\rho, +) \mp i{\cal M}_D(\rho, -)],\nonumber\\
i{\cal M}_D(\rho,  \perp) &=& -i\beta_T^J \sqrt{2} N_1\left[
 \frac{\sqrt \lambda V}{m_{D}+m_{\rho}} \right], \;\;\;
i{\cal M}_D(\rho,||)= -i\beta_T^J\sqrt{2} N_{1} \left[
(m_{D}+m_{\rho})A_1 \right].
\end{eqnarray}

Using the generalized form factor, the  matrix elements for $D$
decays into the spin-0 non-resonating  $\pi\pi$  final state  are
given as
\begin{eqnarray}
 A_0^0 &=& \sqrt{ N_2} i\frac{1}{m_{\pi\pi}} \Bigg[   \frac{\sqrt {\lambda}}{\sqrt{ q^2}}{\cal F}_1^{\pi\pi}(m_{\pi\pi}^2, q^2)  \Bigg] ,\;\;\;
 A_t^0 =\sqrt{ N_2} i \frac{1}{m_{\pi\pi}}\Bigg[ \frac{m_{D}^2-m_{\pi\pi }^2}{\sqrt {q^2}} {\cal F}_0^{\pi\pi}(m_{\pi\pi}^2, q^2)\Bigg],\label{eq:S-waveKpi-formula}
\end{eqnarray}
$N_2=N_1 N_{\rho} \rho_{\pi}/(16\pi^2)$, with $\rho_{\pi}=
\sqrt{1-4m_{\pi}^2/m^2_{\pi\pi}}$.

The above  quantities      can lead to the full angular distributions
\begin{eqnarray}
 \frac{\mathrm{d}^5\Gamma}{\mathrm{d}m_{\pi\pi}^2\mathrm{d}q^2\mathrm{d}\cos\theta_{\pi^+} \mathrm{d}\cos\theta_l \mathrm{d}\phi}
 &=& \frac{3}{8}\Big[I_1(q^2, m_{\pi\pi}^2, \theta_{\pi^+})   \nonumber\\
 && +I_2 (q^2, m_{\pi\pi}^2, \theta_{\pi^+})
 \cos(2\theta_{\ell})  \nonumber\\
 && + I_3(q^2, m_{\pi\pi}^2, \theta_{\pi^+}) \sin^2\theta_{\ell}
 \cos(2\phi) \nonumber\\
 &&+I_4(q^2, m_{\pi\pi}^2, \theta_{\pi^+})  \sin(2\theta_{\ell})\cos\phi \nonumber\\
 && +I_5 (q^2, m_{\pi\pi}^2, \theta_{\pi^+})  \sin(\theta_{\ell}) \cos\phi \nonumber\\
 &&+I_6 (q^2, m_{\pi\pi}^2, \theta_{\pi^+})  \cos\theta_{\ell} \nonumber\\
 && +I_7 (q^2, m_{\pi\pi}^2, \theta_{\pi^+})
 \sin(\theta_{\ell}) \sin\phi\nonumber\\
 && +I_8(q^2, m_{\pi\pi}^2, \theta_{\pi^+})  \sin(2\theta_{\ell})\sin\phi \nonumber\\
 &&+I_9(q^2, m_{\pi\pi}^2, \theta_{\pi^+})  \sin^2\theta_{\ell}
 \sin(2\phi)\Big].
\end{eqnarray}
For the general expressions of $I_i$, we refer the reader to the
appendix and to Refs.~\cite{Pais:1968zza,Lee:1992ih} for the
formulas with the S-, P- and D-waves. In the following, we shall
only consider the S-wave and P-wave contributions and thus the above
general expressions are reduced to:
\begin{eqnarray}
 I_1  &=&
   \frac{1}{4\pi} \left[(1+\hat m_l^2) |A^0_{0}|^2
 +2 \hat m_l^2  |A_t^0|^2\right] + \frac{3}{4\pi} \cos^2\theta_{\pi^{+}} \left[(1+\hat m_l^2) |A^1_{0}|^2
 +2 \hat m_l^2  |A_t^1|^2\right]   \nonumber\\
 &&+ \frac{2\sqrt3 \cos\theta_{\pi^{+}}}{4\pi} \left[  (1+\hat m_l^2) {\rm Re}[A^0_{0} A^{1*}_{0}]   + 2\hat m_l^2 {\rm Re}[A^0_{t} A^{1*}_{t}] \right]
+   \frac{3+\hat m_l^2}{2}  \frac{3}{8\pi}  \sin^2\theta_{\pi^{+}}    [|A^1_{\perp}|^2+|A^1_{||}|^2 ],\nonumber\\
 I_2   &=& -\beta_l     \bigg\{ \frac{1}{4\pi} |A^0_{0}|^2 + \frac{3}{4\pi}\cos^2\theta_{\pi^{+}}    |A^1_{0}|^2   + \frac{2\sqrt3 \cos\theta_{\pi^{+}}}{4\pi}
 {\rm Re}[A^0_{0} A^{1*}_{0}]   \bigg\}+
 \frac{1}{2}\beta_l    \frac{3}{8\pi} \sin^2\theta_{\pi^{+}}    (|A^1_{\perp}|^2+|A^1_{||}|^2),
 \nonumber\\
 I_3&=&
 \beta_l    \frac{3}{8\pi} \sin^2\theta_{\pi^{+}}    (|A^1_{\perp}|^2-|A^1_{||}|^2),
 \nonumber\\
 I_4
  &=&2   \beta_l     \left[ \frac{\sqrt 3 \sin\theta_{\pi^{+}}}{4\sqrt 2\pi} {\rm Re}[A^0_{0}A^{1*}_{||} ] + \frac{3\sin\theta_{\pi^{+}}\cos\theta_{\pi^{+}} }{4\sqrt 2\pi}   {\rm Re}[A^1_{0}A^{1*}_{||} ]  \right],
  \nonumber\\
 I_5
  &=&4\bigg\{\frac{\sqrt 3 \sin\theta_{\pi^{+}}}{4\sqrt 2\pi}  ({\rm Re}[A^0_{0}A^{1*}_{\perp } ] -\hat m_l^2  {\rm Re}[A^0_{t}A^{1*}_{||} ] )      + \frac{  3 \sin\theta_{\pi^{+}}\cos\theta_{\pi^{+}} }{4\sqrt 2\pi} ( {\rm Re}[A^1_{0}A^{1*}_{\perp } ] -\hat m_l^2     {\rm Re}[A^1_{t}A^{1*}_{||} ]) \bigg\} ,
   \nonumber\\
 I_6&=& 4\bigg\{ \frac{3}{8\pi} \sin^2\theta_{\pi^{+}}   {\rm Re}[ A^1_{||}A^{1*}_{\perp} ]  +  \hat m_l^2 \frac{1}{4\pi} {\rm Re}[A_t^0 A_{0}^{0*}] + \hat m_l^2 \frac{3}{4\pi} \cos^2\theta_{\pi^{+}} {\rm Re}[A_t^1 A_{0}^{1*}] \bigg\}
 \nonumber\\
  I_7
 &=& 4\bigg\{\frac{\sqrt 3}{4\sqrt 2\pi} \sin\theta_{\pi^{+}}   ({\rm Im}[A^0_{0}A^{1*}_{||}] - \hat m_l^2 {\rm Im}[ A_t^0 A_\perp^{1*}] )   \nonumber\\
 && +    \frac{3}{4\sqrt 2\pi} \sin\theta_{\pi^{+}} \cos\theta_{\pi^{+}}  ({\rm Im}[A^1_{0}A^{1*}_{||}] - \hat m_l^2 {\rm Im}[ A_t^1 A_\perp^{1*}] )\bigg\}  \nonumber\\
 I_8 &=&
 2 \beta_l  \bigg\{\frac{\sqrt 3}{4\sqrt 2\pi} \sin\theta_{\pi^{+}}    {\rm Im} [A_0^0 A_\perp^{1*}]+ \frac{  3}{4\sqrt 2\pi} \sin\theta_{\pi^{+}} \cos\theta_{\pi^{+}}   {\rm Im} [A_0^1 A_\perp^{1*}]\bigg\},
 \nonumber\\
 I_9
 &=&2 \beta_l  \frac{3}{8\pi} \sin^2\theta_{\pi^{+}} {\rm Im}[A_\perp^1 A_{||}^{1*} ].
 \label{eq:simplified_angularCoefficients}
\end{eqnarray}
Since the phase in P-wave contributions arise from the lineshape
which is the same for different polarizations, the $I_9$ term  and
the second line   in the $I_7$ are zero.

\subsection{Differential and integrated decay widths}

Using the narrow width approximation, we obtain  the integrated
branching fraction:
\begin{eqnarray}
 {\cal B}(D^-\to \rho^0 e^-\bar\nu) &=& (2.24\pm0.09)\times 10^{-3}/(2.16\pm0.36)\times 10^{-3} \mathrm{(LFQM/LCSR)},\\
 {\cal B}(D^-\to \rho^0 \mu^-\bar\nu) &=& (2.15\pm 0.08)\times 10^{-3}/(2.06\pm 0.35)\times 10^{-3} \mathrm{(LFQM/LCSR)},\\
 {\cal B}(\bar D^0\to \rho^+ e^-\bar\nu) &=& (1.73\pm 0.07)\times 10^{-3}/(1.67\pm 0.27)\times 10^{-3}\mathrm{(LFQM/LCSR)},
\end{eqnarray}
where theoretical errors are from the heavy-to-light transition form factors.
These theoretical results are in good agreement with the
data~\cite{Olive:2016xmw}:
\begin{eqnarray}
 {\cal B}(D^-\to \rho^0 e^-\bar\nu) &=& (2.18^{+0.17}_{-0.25})\times 10^{-3},\\
 {\cal B}(D^-\to \rho^0 \mu^-\bar\nu) &=& (2.4\pm0.4)\times 10^{-3},\\
 {\cal B}(\bar D^0\to \rho^+ e^-\bar\nu) &=& (1.77\pm0.16)\times 10^{-3}.
\end{eqnarray}

The starting point for detailed analysis of $D\to \pi\pi\ell\bar\nu$
is to obtain the double-differential distribution
$\mathrm{d}^2\Gamma/\mathrm{d}q^2 \mathrm{d}m_{\pi\pi}^2$ after
performing integration over all the angles
\begin{eqnarray}
\frac{\mathrm{d}^2\Gamma}{\mathrm{d}q^2 \mathrm{d}m_{\pi\pi}^2} &=&
\left(1+\frac{\hat m_l^2}{2}\right)( |A_{0}^0|^2 +  |A_{0}^1|^2  +
|A_{||}^1|^2 + |A_{\perp}^1|^2 )   + \frac{3}{2} \hat m_l^2
(|A_{t}^1|^2 + |A_{t}^0|^2 ),
\end{eqnarray}
where apparently in the massless limit for the involved lepton,  the
total normalization for angular distributions changes to the sum of
the S-wave and P-wave amplitudes
\begin{eqnarray}
\frac{\mathrm{d}^2\Gamma}{\mathrm{d}q^2 \mathrm{d}m_{\pi\pi}^2} =
|A_{0}^0|^2 + |A_{0}^1|^2 + |A_{||}^1|^2 +  |A_{\perp}^1|^2.
\end{eqnarray}

\begin{figure}[h]
\begin{center}
\includegraphics[scale=0.5]{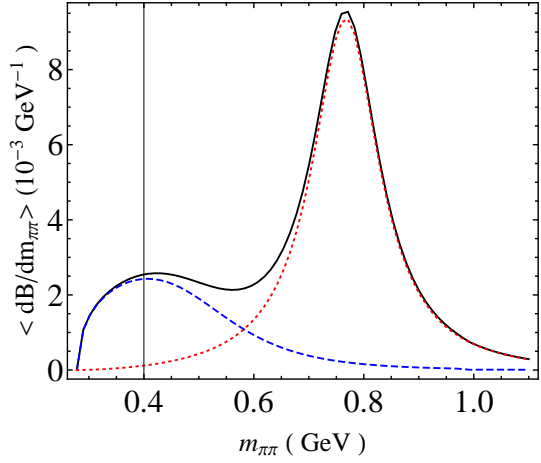}
\caption{ The dependence of branching fraction  on $m_{\pi\pi}$ in
the $D^-\to \pi^+\pi^-e^-\bar\nu_e$ process.  The heavy-to-light
form factors are evaluated by using LCSR } \label{fig:dBdM2pi}
\end{center}
\end{figure}

In  Fig.~\ref{fig:dBdM2pi}, we give
the dependence of branching fraction  on $m_{\pi\pi}$ in the $D^-\to \pi^+\pi^-e^-\bar\nu_e$ process.    The solid, dashed, and dotted    curves correspond to the total,  S-wave and P-wave contributions.
For the S-wave contribution, there is no resonance around $0.98$ GeV, and theoretically, this should be a dip.

Due to the quantum number constraint, the process $\bar D^0\to
\pi^+\pi^0 \ell\bar\nu$ receives only a P-wave contribution and  $
D^-\to \pi^0\pi^0 \ell\bar\nu$ is generated by the S-wave term.

\begin{figure}[h]
\begin{center}
\includegraphics[scale=0.5]{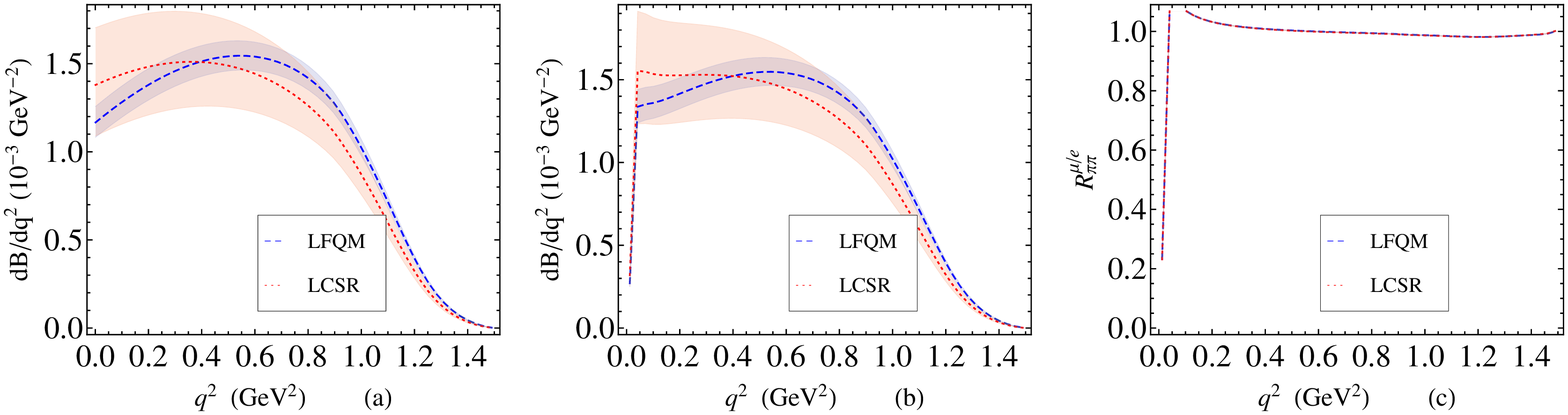}
\caption{ Differential decay widths    for the  $D^-\to
\pi^+\pi^-\ell \bar\nu_{\ell}$ with $\ell=e$ in panel (a) and $\ell=
\mu$ in panel (b). The $q^2$-dependent ratio $R_{\pi\pi}^{\mu/e}$ as
defined in Eq.~\eqref{eq:R2piMuOverE} is given in panel (c). The
\textit{dashed} and \textit{dotted curves} are produced using the
LFQM and LCSR results for $D\to \rho$ form factors.  Errors from the
form factors are shown as \textit{shadowed bands}, and  most errors
cancel in the ratio $R_{\pi\pi}^{\mu/e}$ given in panel (c)   }
\label{fig:dBdq2D2Rho}
\end{center}
\end{figure}

To match the kinematics constraints  implemented in experimental
measurements, one may explore the generic observable  with
$m_{\pi\pi}^2$ integrated out:
\begin{eqnarray}
 \langle O\rangle = \int_{(m_{\rho}-\delta_m)^2}^{(m_{\rho}+\delta_m)^2} \mathrm{d}m^2_{\pi\pi} \frac{\mathrm{d}O}{\mathrm{d}m_{\pi\pi}^2}.
\end{eqnarray}
We use the following choice  in
our study of $D\to \pi\pi \ell\bar\nu$:
\begin{eqnarray}
 \delta_m = \Gamma_\rho.
\end{eqnarray}
In the narrow width-limit, the integration of the lineshape is conducted as
\begin{eqnarray}
 \int \mathrm{d}m^2_{\pi\pi}|L_{\rho}(m^2_{\pi\pi})|^2 = {\cal B}(\rho^0\to \pi^-\pi^+)=1.
\end{eqnarray}
However, with the explicit form given in
Eq.~\eqref{eq:KstarLineShape}, we find  that the integration
\begin{eqnarray}
 \int_{(m_{\rho}-\delta_m)^2}^{(m_{\rho}+\delta_m)^2} \mathrm{d}m^2_{\pi\pi}|L_{\rho}(m^2_{\pi\pi})|^2 =0.70
\end{eqnarray}
is below the expected value.
On the other hand,  the integrated  S-wave lineshape in this region is
\begin{eqnarray}
\int_{(m_{\rho}-\delta_m)^2}^{(m_{\rho}+\delta_m)^2}
\mathrm{d}m^2_{\pi\pi}|L_{S}(m^2_{\pi\pi})|^2 ={  0.37},
\end{eqnarray}
which is smaller but  at the same order. Integrated from
$m_{\rho}-\Gamma_{\rho}$ to $m_{\rho}+\Gamma_{\rho}$, we have
\begin{eqnarray}
 {\cal B}(D^-\to \rho^0 (\to \pi^+\pi^-) e^-\bar\nu) = (1.57\pm 0.07)\times 10^{-3}/(1.51\pm 0.26)\times 10^{-3}\ (\mathrm{LFQM/LCSR}),\\
 {\cal B}(D^-\to \rho^0 (\to \pi^+\pi^-) \mu^-\bar\nu) = (1.57\pm 0.07)\times 10^{-3}/(1.51\pm 0.26)\times 10^{-3}\ (\mathrm{LFQM/LCSR}).
\end{eqnarray}
The S-wave branching fractions for $2m_{\pi}<m_{\pi\pi}<
1.0~\mathrm{GeV}$ are given as
\begin{eqnarray}
 {\cal B}(D^-\to    (\pi^+\pi^-)_S e^-\bar\nu) =   (6.99\pm 2.46)\times 10^{-4},\\
 {\cal B}(D^-\to  (  \pi^+\pi^-)_S \mu^-\bar\nu) =  (7.20\pm 2.52)\times 10^{-4}.
\end{eqnarray}
Above 1 GeV, the unitarized $\chi$PT will fail and thus we lack any
reliable prediction.

 Furthermore, one may   explore the $q^2$-dependent ratio
\begin{eqnarray}
\label{eq20:DDRRDst2} R_{\pi\pi}^{\mu/e}(q^2) &=&\frac{\langle
\mathrm{d}\Gamma(D\to \pi\pi \mu\bar\nu_{\mu})/\mathrm{d}q^2 \rangle
}{\langle \mathrm{d}\Gamma( D\to \pi\pi
e\bar\nu_e)/\mathrm{d}q^2\rangle }.\label{eq:R2piMuOverE}
\end{eqnarray}
Differential decay widths for  $D\to \pi\pi\ell \bar\nu_{\ell}$ are
given in Fig.~\ref{fig:dBdq2D2Rho},  with $\ell= e$ in panel (a) and
$\ell=\mu$ in panel (b). The $q^2$-dependent ratio
$R_{\pi\pi}^{\mu/e}$  is given in panel (c). Errors from the form
factors and QCD condensate parameter $B_0$ are shown as shadowed
bands, and most errors cancel in the ratio $R_{\pi\pi}^{\mu/e}$
given in panel (c).

 \subsection{Distribution in $\theta_{\pi^{+}}$ }

\begin{figure}[h]
\begin{center}
\includegraphics[scale=0.5]{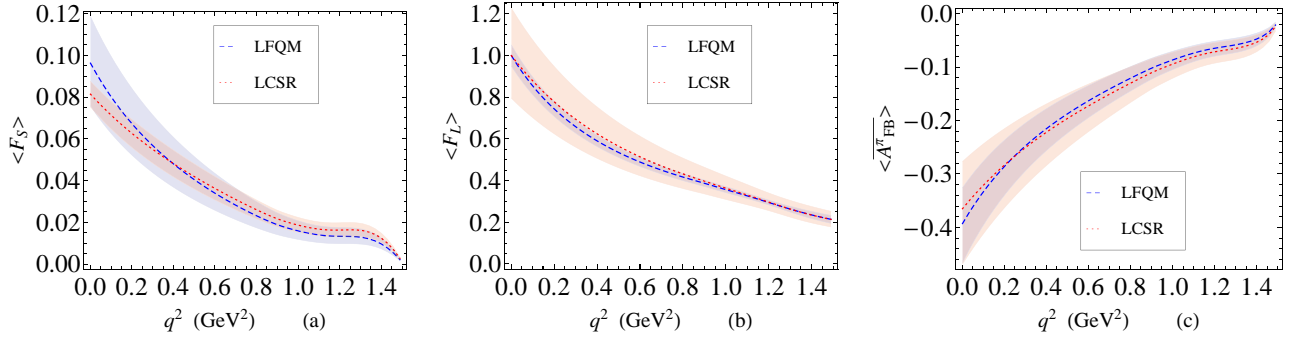}
\caption{Same as Fig.~\ref{fig:dBdq2D2Rho} but for the S-wave
contributions (a) and the longitudinal  polarizations in P-wave
contributions (b)  to the $D\to \pi\pi \ell\bar\nu_{\ell}$, and the
forward--backward asymmetry $\overline {A_{FB}^{\pi}}$ (c). Notice
that, for the  $\overline {A_{FB}^{\pi}}$, there is a sign ambiguity
arising from the use of Watson theorem.  These diagrams are for the
light lepton $e$, while the results  for the $\mu$ lepton are
similar} \label{fig:FSBsKstar}
\end{center}
\end{figure}

We explore the distribution in $\theta_{\pi^{+}}$:
\begin{eqnarray}
\frac{\mathrm{d}^3\Gamma} { \mathrm{d}q^2 \mathrm{d}m_{\pi\pi}^2 \mathrm{d}\cos\theta_{\pi^{+}}}  &=&  \frac{\pi}{2} (3I_1-I_2) \nonumber\\
&=& \frac{1}{8} \bigg\{ (4+2\hat m_l^2) |A_0^0|^2 + 6\hat m_l^2 |A_t^0|^2 \nonumber\\
&&  + \sqrt{3} (8+4\hat m_l^2) \cos\theta_{\pi^{+}} {\rm Re}[A_0^0 A_{0}^{1*}] + 12\sqrt{3} \hat m_l^2 \cos\theta_{\pi^{+}}  {\rm Re}[A_t^0 A_{t}^{1*}] \nonumber\\
&&+ (12+ 6\hat m_l^2) |A_0^1|^2 \cos^2\theta_{\pi^{+}} + 18\hat m_l^2 \cos^2\theta_{\pi^{+}} |A_t^1|^2   \nonumber\\
&&  + (6+3\hat m_l^2) \sin^2\theta_{\pi^{+}}  (|A_\perp^1|^2 +
|A_{||}^1|^2)  \bigg\}.\label{eq:theta_K_distribution}
\end{eqnarray}

Compared to the distribution with only P-wave contribution, namely
$D\to \rho(\to \pi\pi)\ell\bar\nu$,  the first two lines   of
Eq.~\eqref{eq:theta_K_distribution} are new: the first one is the
S-wave $\pi\pi$ contribution,  while the second line  arises from
the interference of   S-wave and P-wave.  Based on this
interference,  one can define a forward--backward  asymmetry for the
involved pion,
\begin{eqnarray}
 A_{FB}^{\pi} &\equiv &   \bigg[\int_{0}^1
 - \int_{-1}^0\bigg] \mathrm{d}\cos\theta_{\pi^{+}} \frac{\mathrm{d}^3\Gamma}{ \mathrm{d}q^2 \mathrm{d}m_{\pi\pi}^2
  \mathrm{d}\cos\theta_{\pi^{+}}}  \nonumber\\
 &=&  \frac{\sqrt{3}}{2}  (2+ \hat m_l^2)  {\rm Re}[A_0^0 A_{0}^{1*}] + \frac{3\sqrt{3}}{2}  \hat m_l^2   {\rm Re}[A_t^0 A_{t}^{1*}].
\end{eqnarray}


We define the polarization fraction  at a given value of $q^2$ and
$m_{\pi\pi}^2$:
\begin{eqnarray}
 {\cal F}_{S} (q^2, m_{\pi\pi}^2)  = \frac{ (1+\hat m_l^2/2) |A_0^0|^2 + 3/2 \hat m_l^2 |A_t^0|^2 } {  {\mathrm{d}^2\Gamma}/({\mathrm{d}q^2 \mathrm{d}m_{\pi\pi}^2})} ,\nonumber\\
 {\cal F}_{P } (q^2, m_{\pi\pi}^2)  = \frac{  (1+\hat m_l^2/2)(|A_{0}^1|^2  + |A_{||}^1|^2 +  |A_{\perp}^1|^2 ) + 3/2\hat m_l^2 |A^1_t|^2 } { {\mathrm{d}^2\Gamma}/({\mathrm{d}q^2 \mathrm{d}m_{\pi\pi}^2})} ,
\end{eqnarray}
and also
\begin{eqnarray}
 F_L (q^2, m_{\pi\pi}^2)  = \frac{   (1+\hat m_l^2/2)|A_{0}^1(q^2, m_{\pi\pi}^2)|^2  + 3/2\hat m_l^2 |A^1_t|^2  } {   (1+\hat m_l^2/2)(|A_{0}^1|^2  + |A_{||}^1|^2 +  |A_{\perp}^1|^2 ) + 3/2\hat m_l^2 |A^1_t|^2 },\nonumber\\
 \overline {A_{FB}^{\pi}}(q^2, m_{\pi\pi}^2)=  \frac{   {\sqrt{3}}/{2}  (2+\hat m_l^2)  {\rm Re}[A_0^0 A_{0}^{1*}] +  {3\sqrt{3}}/{2}  \hat m_l^2   {\rm Re}[A_t^0 A_{t}^{1*}] } { {\mathrm{d}^2\Gamma}/({\mathrm{d}q^2 \mathrm{d}m_{\pi\pi}^2})} .
\end{eqnarray}
By definition, ${\cal F}_S+{\cal F}_P=1$.

In Fig.~\ref{fig:FSBsKstar},{ we give our results for the S-wave
fraction $\langle  F_S\rangle$ (panel (a)), longitudinal
polarization fraction $\langle  F_L\rangle$  in P-wave contributions
(panel (b)) and the asymmetry $\langle  \overline {A_{FB}^{\pi}}
\rangle$ (panel (c)). Only  the curves  for the light lepton $e$ are
shown since the results for the $\mu$ lepton are similar. }These
observables and the following ones are defined by the integration
over $m_{\pi\pi}^2$; for instance,
\begin{eqnarray}
 \langle  F_{S} (q^2) \rangle = \frac{\int \mathrm{d}m_{\pi\pi}^2 [ (1+\hat m_l^2/2) |A_0^0|^2 + 3/2 \hat m_l^2 |A_t^0|^2]  } {\int \mathrm{d}m_{\pi\pi}^2  {\mathrm{d}^2\Gamma}/({\mathrm{d}q^2 \mathrm{d}m_{\pi\pi}^2})} ,
\end{eqnarray}
and likewise for the others.

\subsection{ Distribution in $\theta_l$ and forward--backward asymmetry}

\begin{figure}
\begin{center}
\includegraphics[scale=0.5]{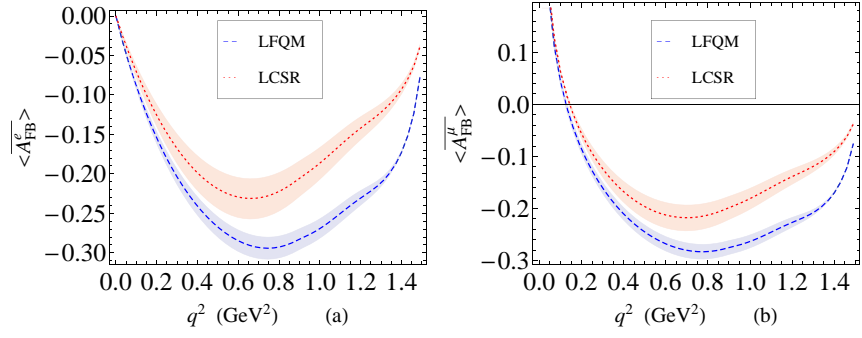}
\caption{Same as Fig.~\ref{fig:dBdq2D2Rho} but for the asymmetry
$\overline {{\cal A}_{FB}^l} $   in the $D\to \pi\pi
\ell\bar\nu_{\ell}$} \label{fig:AFBD2Rho}
\end{center}
\end{figure}

Integrating over $\theta_{\pi^{+}}$ and $\phi$, we have the
distribution:
\begin{eqnarray}
\frac{\mathrm{d}^3\Gamma} { \mathrm{d}q^2 \mathrm{d}m_{\pi\pi}^2 \mathrm{d}\cos\theta_l}  &=& \frac{3\pi}{4} \int \mathrm{d}\cos\theta_{\pi^{+}} (I_1 +I_2\cos(2\theta_l) +I_6 \cos\theta_l )\nonumber\\
&=& \frac{3}{4} \hat m_l^2 ( (|A_{t}^0|^2 +|A_{t}^1|^2 ))  + \frac{3}{2} \cos\theta_l  ( {\rm Re}[A_{||}^1 A_{\perp}^{1*}] + \hat m_l^2  {\rm Re}[A_{t}^0 A_{0}^{0*}+A_{t}^1 A_{0}^{1*}] )
\nonumber\\
&& +\frac{3}{4} [1 -(1-\hat m_l^2)\cos^2\theta_l] (|A_{0}^0|^2 +|A_{0}^1|^2 ) + \frac{3}{8} [ (1+\hat m_l^2) + (1-\hat m_l^2) \cos^2\theta_l ]  (|A_{||}^1|^2 +|A_{\perp}^1|^2 ).
\end{eqnarray}
The forward--backward asymmetry is defined as
\begin{eqnarray}
 {\cal A}_{FB}^{l}  &\equiv &   \bigg[\int_{0}^1
 - \int_{-1}^0\bigg] \mathrm{d}\cos\theta_l \frac{\mathrm{d}^3\Gamma}{ \mathrm{d}q^2 \mathrm{d}m_{\pi\pi}^2
  \mathrm{d}\cos\theta_l }  =  \frac{3}{2}   ( {\rm Re}[A_{||}^1 A_{\perp}^{1*}] + \hat m_l^2  {\rm Re}[A_{t}^0 A_{0}^{0*}+A_{t}^1 A_{0}^{1*}] ) ,
\end{eqnarray}
and the results  for $\overline {{\cal A}_{FB}^l} $  are given in
Fig.~\ref{fig:AFBD2Rho}.

 \subsection{Distribution in the azimuth  angle  $\phi$ }

\begin{figure}
\begin{center}
\includegraphics[scale=0.5]{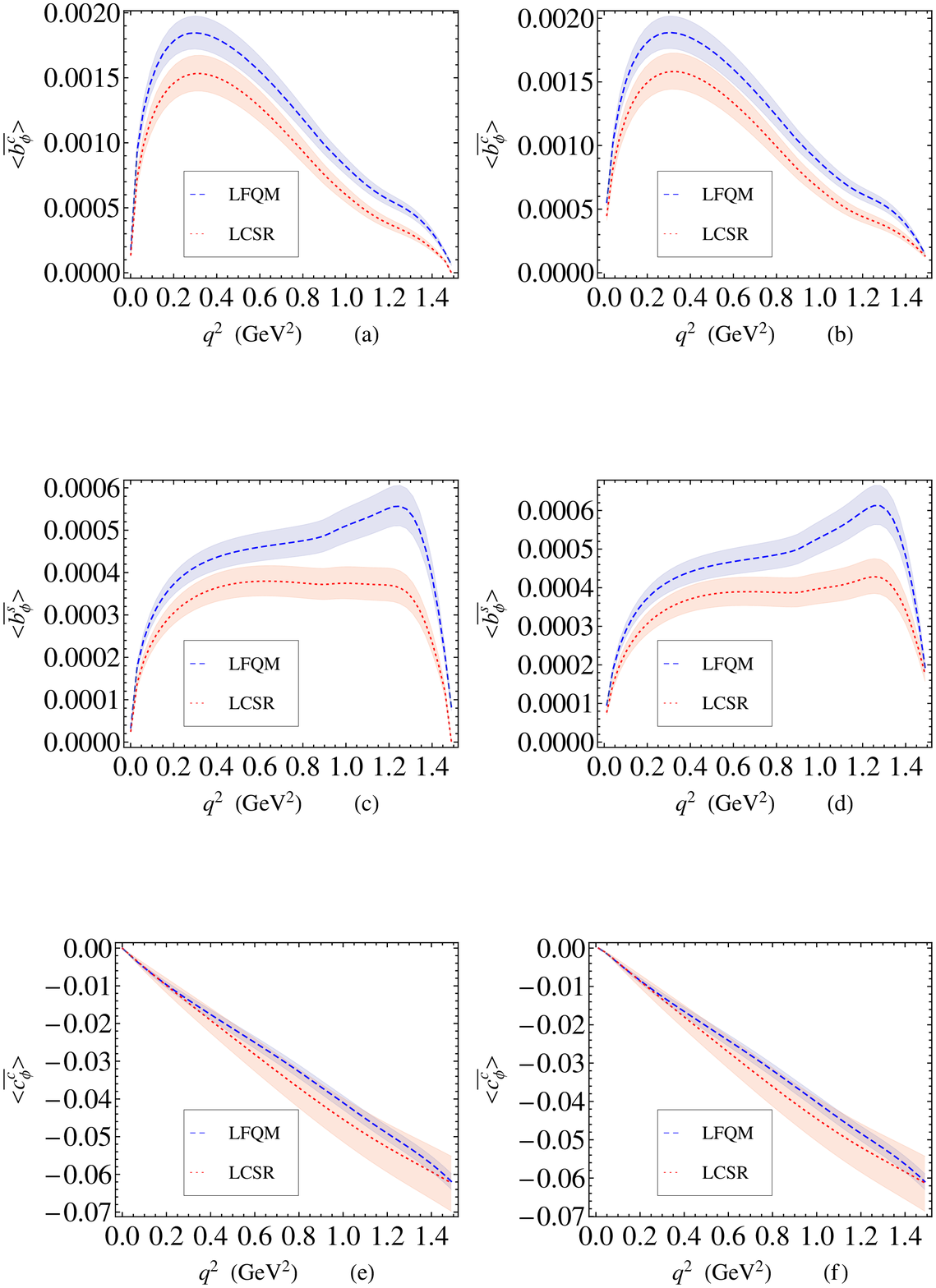}
\caption{Same as Fig.~\ref{fig:dBdq2D2Rho} but for the normalized
coefficients in the $\phi$ distributions of the $D^-\to \pi^+ \pi^-
\ell\bar\nu_{\ell}$.  The \textit{left panels} (a, c, e) are for the
light lepton $e$, while the \textit{right panels} (b, d, f) are for
the $\mu$ lepton } \label{fig:bcphiD2Rho}
\end{center}
\end{figure}

The angular distribution in $\phi$ is derived  as
\begin{eqnarray}
\frac{\mathrm{d}^3\Gamma} { \mathrm{d}q^2 \mathrm{d}m_{\pi\pi}^2
\mathrm{d}\phi} &=& a_\phi +b_\phi^c \cos\phi + b_\phi^s \sin\phi +
c_\phi^c \cos(2\phi) + c_\phi^s \sin(2\phi)
\end{eqnarray}
with
\begin{eqnarray}
 a_\phi &=& \frac{1}{2\pi} \frac{\mathrm{d}^2\Gamma} { \mathrm{d}q^2 \mathrm{d}m_{\pi\pi}^2}  ,\nonumber\\
 b_\phi^c &=& \frac{3}{16}\pi \int  I_5 \mathrm{d}\cos\theta_{\pi^{+}} =  \frac{3\sqrt3}{32\sqrt 2\pi} \big({\rm Re}[A_0^0 A_{\perp}^{1*}]-\hat m_l^2 {\rm Re}[A_t^0 A_{\perp}^{1*}]\big)\nonumber\\
 b_\phi^s &=& \frac{3}{16}\pi \int  I_7  \mathrm{d}\cos\theta_{\pi^{+}}=  \frac{3\sqrt3}{32\sqrt 2\pi} \big({\rm Im}[A_0^0 A_{\perp}^{1*}]-\hat m_l^2 {\rm Im}[A_t^0 A_{\perp}^{1*}]\big)\nonumber\\
 c_\phi^c &=& \frac{1}{2}   \int I_3 \mathrm{d}\cos\theta_{\pi^{+}} =  \frac{1}{4\pi} \beta_l (|A_{\perp}^1|^2-|A_{||}^1|^2), \;\;\;
 c_\phi^s =\frac{1}{2}   \int  I_9 \mathrm{d}\cos\theta_{\pi^{+}}=  \frac{1}{ 2 \pi } \beta_l {\rm Im}[A_{\perp}^1 A_{||}^{1*}].
\end{eqnarray}
Since the complex phase in the P-wave amplitudes comes from the
Breit--Wigner lineshape, the coefficient $c_\phi^s$ vanishes.

Numerical results for the normalized coefficients  using the two
sets of form factors   are shown in Fig.~\ref{fig:bcphiD2Rho}. The
coefficients $ b_\phi^c$ and $ b_\phi^s$ contain a very small
prefactor, $ {3\sqrt3}/({32\sqrt 2\pi} )\sim 0.037$, and thus are
numerically tiny, as shown in this figure. The $ c_\phi^c$ is also
small due to the cancellation between the $|A_{\perp}|^2$ and
$|A_{||}|^2$.



\subsection{Polarization of $\mu$ lepton}

\begin{figure}
\begin{center}
\includegraphics[scale=0.6]{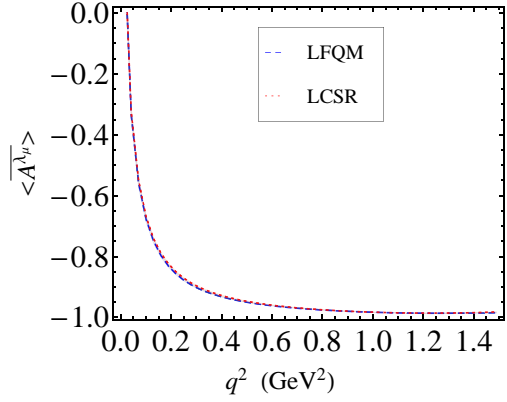}
\caption{Same as Fig.~\ref{fig:dBdq2D2Rho} but for the polarization
distribution  of $D\to \pi\pi \mu\bar\nu_{\mu}$. Theoretical errors
are negligible } \label{fig:AlambdaD2Rho}
\end{center}
\end{figure}

In this work, we also give the polarized angular distributions as
\begin{eqnarray}
 \frac{\mathrm{d}^5\Gamma(\lambda_\mu)}{\mathrm{d}m_{\pi\pi}^2\mathrm{d}q^2\mathrm{d}\cos\theta_{\pi^{+}} \mathrm{d}\cos\theta_l \mathrm{d}\phi}
 &=& \frac{3}{8}\Big[I_1^{(\lambda_\mu)}   +I_2 ^{(\lambda_\mu)}
 \cos(2\theta_l)   + I_3^{(\lambda_\mu)} \sin^2\theta_l
 \cos(2\phi) \nonumber\\
 &&+I_4^{(\lambda_\mu)}  \sin(2\theta_l)\cos\phi  +I_5^{(\lambda_\mu)} \sin(\theta_l) \cos\phi  +I_6 ^{(\lambda_\mu)}  \cos\theta_l \nonumber\\
 && +I_7^{(\lambda_\mu)}
 \sin(\theta_l) \sin\phi  +I_8^{(\lambda_\mu)}  \sin(2\theta_l)\sin\phi  +I_9^{(\lambda_\mu)} \sin^2\theta_l
 \sin(2\phi)\Big],
\end{eqnarray}
with the coefficients
\begin{eqnarray}
 I_1^{(-1/2)}    &=&  |A_{0}|^2  +\frac{3}{2} (|A_{\perp}|^2
 +|A_{||}|^2), \;\;\;\;
 I_2^{(-1/2)}   = -  |A_{0}|^2+ \frac{1}{2}  (|A_{\perp}|^2
 +|A_{||}|^2),
 \nonumber\\
  I_3^{(-1/2)}    &=&   |A_{\perp}|^2-|A_{||}|^2 ,\;\;\;
 I_4^{(-1/2)}    =  2{\rm Re}(A_{0}A_{||}^*),\;\;\;
 I_5^{(-1/2)}
  =4 {\rm Re}(A_{0}A_{\perp}^*), \nonumber\\
 I_6^{(-1/2)}    & =&  4
  {\rm Re}(A_{||}A^*_{\perp}),\;\;\;\;
   I_7^{(-1/2)}   = 4 {\rm Im}(A_{0}A^*_{||})  , \;\;\;
 I_8^{(-1/2)}     =  2   {\rm Im}(A_{0}A^*_{\perp}), \;\;\;\;
 I_9^{(-1/2)}  =2    {\rm Im}(A_{\perp }A^*_{||}).\label{eq:polarisedAngularCoefficients}
\end{eqnarray}
The coefficients for the $\lambda_\mu=1/2$ are easily obtained by
comparing Eqs.~\eqref{eq:polarisedAngularCoefficients} and
\eqref{eq:simplified_angularCoefficients}. For instance, the lepton
polarization fraction is defined as
\begin{eqnarray}
 \overline  {A^{\lambda_\mu}}(q^2, m_{\pi\pi}^2 ) &=&  \frac{ {\mathrm{d}^2\Gamma^{(1/2)}}/{\mathrm{d}q^2\mathrm{d}m_{\pi\pi}^2 }- {\mathrm{d}^2\Gamma^{(-1/2)}}/{\mathrm{d}q^2\mathrm{d}m_{\pi\pi}^2 }}{ {\mathrm{d}^2\Gamma}/{\mathrm{d}q^2\mathrm{d}m_{\pi\pi}^2 }} \nonumber\\
  &=&  \frac{ \big(-1+ {\hat m_l^2}/{2}\big)( |A_{0}^0|^2 +  |A_{0}^1|^2  + |A_{||}^1|^2 +  |A_{\perp}^1|^2 )   + \frac{3}{2} \hat m_l^2 (|A_{t}^1|^2 + |A_{t}^0|^2 )}{ {\mathrm{d}^2\Gamma}/{\mathrm{d}q^2\mathrm{d}m_{\pi\pi}^2 }},
\end{eqnarray}
and we show the numerical results in   Fig.~\ref{fig:AlambdaD2Rho}.

\subsection{Theoretical uncertainties}

Before closing this section, we will briefly  discuss the
theoretical uncertainties in this analysis.  The parametric errors
in  heavy-to-light transition form factors and   QCD condensate
parameter $B_0$ have been included in the above. As one can see,
these uncertainties are sizable to  branching fractions and other
related observables, but are negligible in the ratios like
$R_{\pi\pi}^{\mu/e}$. This is understandable, since most
uncertainties will cancel in the ratio.

For the heavy-to-light form factors, we have used the LCSR and LFQM
results. In LCSR,  the theoretical accuracy  for most form factors
is at leading order in $\alpha_s$. An analysis of $B_s\to
f_0$~\cite{Colangelo:2010bg} has indicated the NLO radiative
corrections to form factors may reach $20\%$. The radiative
corrections are, in general, channel-dependent but  should be
calculated in a high precision study. It should be pointed out that
radiative corrections in the light-front quark model is not
controllable.

A third type of uncertainties resides in the scalar $\pi\pi$ form
factor. In this work, we have used the unitarized results from
Refs.~\cite{Oller:1997ti,Lahde:2006wr}, where the low-energy
constants ($L_i$s) are obtained by fitting the $J/\psi$ decay data.
A Muskhelishvili--Omn\`{e}s formalism has been developed for the
scalar $\pi\pi$ form factor in Ref.~\cite{Daub:2015xja}. Compared to
the results in Ref.~\cite{Daub:2015xja}, we find  an overall
agreement in  the shape of the non-strange $\pi\pi$ form factor, but
the modulus from Ref.~\cite{Daub:2015xja} is about $20\%$ larger.
This would  induce  about $40\%$ uncertainties to the branching
ratios of the $D\to \pi\pi\ell\bar\nu_\ell$, while the results for
the ratio observables are not affected.

Finally, the Watson theorem does not always  guarantee the use of
Eq.~\eqref{eq:matching}, the matching of $D\to \pi\pi$ form factor
and $D\to f_0$ form factors. As we have discussed in Sec.~II, such
an approximation might be improved in the future.

\section{Conclusions}
\label{sec:conclusions}

In summary, we have presented a theoretical analysis of  the $D^-\to
\pi^+\pi^- \ell\bar\nu$ and  $\bar D^0\to \pi^+\pi^0 \ell\bar\nu$
decays. We have   constructed  a general  angular distribution which
can include arbitrary partial waves of $\pi\pi$. Retaining the
S-wave and P-wave contributions we have  studied  the branching
ratios, forward--backward asymmetries and a few other observables.
The P-wave contribution is dominated by $\rho^0$ resonance, and the
S-wave contribution is analyzed using the unitarized chiral
perturbation theory.  The obtained branching fraction for $D\to
\rho\ell\nu$,  at the order $10^{-3}$, is consistent with the
available experimental data, while  the  S-wave contribution is
found to have  a branching ratio  at the order of $10^{-4}$, and
this prediction can be tested  by experiments like BESIII and LHCb.
The BESIII collaboration  has accumulated about $10^7$ events of the
$D^0$ and will collect about  3 fb$^{-1}$ data  at the
center-of-mass $\sqrt s= 4.17$ GeV to produce the
$D_s^+D_s^-$~\cite{Ablikim:2014cea,Asner:2008nq}. All these data can
be used to study the charm decays into the $f_0$ mesons. In
addition, sizable branching fractions  also indicate a promising
prospect at the ongoing  LHC experiment~\cite{Bediaga:2012py}, the
forthcoming Super-KEKB factory~\cite{Aushev:2010bq} and the
under-design Super Tau-Charm factory. Future measurements can   be
used to study the $\pi$--$\pi$ scattering phase shift.

\section*{Acknowledgements}
We thank Jian-Ping Dai,  Liao-Yuan Dong, Hai-Bo Li and Lei Zhang for useful discussions.
This work is supported  in part   by National  Natural
Science Foundation of China under Grant
 Nos.11575110, 11655002,  Natural  Science Foundation of Shanghai under Grant  No.
15DZ2272100 and No. 15ZR1423100,  by the Young Thousand Talents Plan,   and  by    Key Laboratory for Particle Physics, Astrophysics and Cosmology, Ministry of Education.

\begin{appendix}
\section{Angular coefficients}

In the angular distribution, the coefficients have the form
\begin{eqnarray}
 I_1   &=& (1+\hat m_l^2) |A_{0}|^2
 +2 \hat m_l^2  |A_t|^2 + (3+\hat m_l^2)/2(|A_{\perp}|^2
 +|A_{||}|^2)
 \nonumber\\
 I_2   &=& -\beta_l   |A_{0}|^2+  \beta_l /2 (|A_{\perp}|^2
 +|A_{||}|^2), \;\;\;\;\;\;\;\;\;\;
  I_3    = \beta_l (|A_{\perp}|^2-|A_{||}|^2),\;\;\;\;\;\;\;\;\;\;\;
 I_4    =  2 \beta_l  [{\rm Re}(A_{0}A_{||}^*)],\nonumber\\
 I_5   &=&4  [{\rm Re}(A_{0}A_{\perp}^*) -\hat m_l^2 {\rm Re}(A_{t}A_{||}^*) ],\;\;\;\;\;\;\;\;\;\;\;\;\;
 I_6     =  4  [{\rm Re}(A_{||}A^*_{\perp})+ \hat m_l^2 {\rm Re}(A_{t}A^*_{0})],\nonumber\\
 I_7   &=& 4[{\rm Im}(A_{0}A^*_{||})-\hat m_l^2 {\rm Im}(A_{t}A^*_{\perp})],\;\;\;\;\;\;\;\;\;\;\;\;\;
 I_8   =  2 \beta_l   [{\rm Im}(A_{0}A^*_{\perp})],\;\;\;\;\;\;\;\;\;\;\;\;\;\;\;
 I_9   = 2\beta_l   [{\rm Im}(A_{\perp }A^*_{||})]. \label{eq:angularCoefficients}
\end{eqnarray}
Substituting the  expressions for $A_i$ into the above equation, we obtain the general  expressions
\begin{eqnarray}
 I_1(q^2, m_{\pi\pi}^2, \theta_{\pi^{+}})  &=&
 \sum_{J=0,...}  \bigg\{ |Y_J^0(\theta_{\pi^{+}}, 0)|^2 \left[(1+\hat m_l^2) |A^J_{0}|^2
 +2 \hat m_l^2  |A_t^J|^2\right]   \nonumber\\
 &&+ 2\sum_{ J'=J+1, ...} Y_J^0(\theta_{\pi^{+}}, 0)Y_{J'}^0(\theta_{\pi^{+}}, 0) \left[ \cos(\delta_{0}^J -
 \delta^{J'}_{0})|A^J_{0}||A^{J'*}_{0}|   + 2\hat m_l^2
 \cos (\delta_{t}^J -\delta_{t}^{J'} )|A^J_t||A^{J'}_t|\right]\bigg\}
 \nonumber\\
 & &  +  \frac{3+\hat m_l^2}{2}  \sum_{J=1,...}  \bigg\{ |Y_J^{-1}(\theta_{\pi^{+}}, 0)|^2 \left[   [|A^J_{\perp}|^2+|A^J_{||}|^2 ]  \right] \nonumber\\
 && +  \sum_{ J'=J+1, ...} Y_J^{-1}(\theta_{\pi^{+}}, 0)Y_{J'}^{-1}(\theta_{\pi^{+}}, 0) \left[    2\cos(\delta_{\perp}^{J}
 - \delta_{\perp}^{J'})|A_{\perp}^J||A_{\perp}^{J'}| \right]  \bigg\},
  \end{eqnarray}
  \begin{eqnarray}
 I_2(q^2, m_{\pi\pi}^2, \theta_{\pi^{+}})   &=& -\beta_l   \sum_{J=0,...}  \bigg\{ |Y_J^0|^2   |A^J_{0}(\theta_{\pi^{+}}, 0)|^2   + 2\sum_{ J'=J+1, ...} Y_J^0(\theta_{\pi^{+}}, 0)Y_{J'}^0(\theta_{\pi^{+}}, 0)
  \cos(\delta_{0}^J - \delta^{J'}_{0})|A^J_{0} A^{J'}_{0}|  \bigg\} \nonumber\\
  & &+
 \frac{1}{2}\beta_l  \sum_{J=1,...}  \bigg\{  |Y_J^{-1}(\theta_{\pi^{+}}, 0)|^2  (|A^J_{\perp}|^2+|A^J_{||}|^2) \nonumber\\
 &&       +2\sum_{ J'=J+1} Y_J^{-1}(\theta_{\pi^{+}}, 0)Y_{J'}^{-1}(\theta_{\pi^{+}}, 0)\left[ \cos(\delta_{\perp}^J
 - \delta^{J'}_{\perp})|A^J_{\perp} A^{J'}_{\perp}| + \cos(\delta_{||}^J
 - \delta^{J'}_{||})|A^J_{||}A^{J'}_{||}| \right] \bigg\},
  \end{eqnarray}
  \begin{eqnarray}
 I_3(q^2, m_{\pi\pi}^2, \theta_{\pi^{+}})  &=& \beta_l \sum_{J=1,...}  \bigg\{  |Y_J^{-1}(\theta_{\pi^{+}}, 0)|^2  (|A^J_{\perp}|^2-|A^J_{||}|^2)         \nonumber\\
 && +2\sum_{ J'=J+1,...} Y_J^{-1}(\theta_{\pi^{+}}, 0)Y_{J'}^{-1}(\theta_{\pi^{+}}, 0)\left[ \cos(\delta_{\perp}^J
 - \delta^{J'}_{\perp})|A^J_{\perp} A^{J'}_{\perp}| - \cos(\delta_{||}^J
 - \delta^{J'}_{||})|A^J_{||}A^{J'}_{||}| \right] \bigg\},
  \end{eqnarray}
  \begin{eqnarray}
 I_4(q^2, m_{\pi\pi}^2, \theta_{\pi^{+}})
  &=&2 \beta_l   \sum_{J=0, ...} \sum_{J'=1, ..} \left[  Y_J^0(\theta_{\pi^{+}}, 0) Y_{J'}^{-1}(\theta_{\pi^{+}}, 0) | A^J_{0}A^{J'*}_{||} | \cos(\delta_{0}^J -\delta_{||}^{J'})  \right],
  \end{eqnarray}
  \begin{eqnarray}  
 I_5(q^2, m_{\pi\pi}^2, \theta_{\pi^{+}})
  &=&4   \sum_{J=0, ...} \sum_{J'=1, ..}Y_J^0(\theta_{\pi^{+}}, 0) Y_{J'}^{-1} (\theta_{\pi^{+}}, 0) \left[   | A^J_{0}A^{J'*}_{\perp} | \cos(\delta_{0}^J -\delta_{\perp}^{J'}) -\hat m_l^2  | A^J_{t}A^{J'*}_{||} | \cos(\delta_{t}^J -\delta_{||}^{J'})\right],
  \end{eqnarray}
  \begin{eqnarray}
 I_6(q^2, m_{\pi\pi}^2, \theta_{\pi^{+}})  &=& 4 \sum_{J,J'=1,...}  \bigg\{   Y_J^{-1} (\theta_{\pi^{+}}, 0)Y_{J'}^{-1} (\theta_{\pi^{+}}, 0)| A^J_{||}A^{J'*}_{\perp} | \cos(\delta_{||}^J -\delta_{\perp}^{J'})  \bigg\}   \nonumber\\
 && +  \hat m_l^2 \sum_{J,J'=0,...}  \bigg\{   Y_J^{0}(\theta_{\pi^{+}}, 0) Y_{J'}^{0}(\theta_{\pi^{+}}, 0) | A^J_{t}A^{J'*}_{0} | \cos(\delta_{t}^J -\delta_{0}^{J'})  \bigg\},
  \end{eqnarray}
  \begin{eqnarray}
  I_7(q^2, m_{\pi\pi}^2, \theta_{\pi^{+}})
 &=&4 \sum_{J=0, ...} \sum_{J'=1, ..} Y_J^0(\theta_{\pi^{+}}, 0) Y_{J'}^{-1}(\theta_{\pi^{+}}, 0)  \left[  | A^J_{0}A^{J'*}_{||} | \sin(\delta_{0}^J -\delta_{||}^{J'}) -\hat m_l^2 | A^J_{t}A^{J'*}_{\perp} | \sin(\delta_{t}^J -\delta_{\perp}^{J'})\right],
  \end{eqnarray}
  \begin{eqnarray}
 I_8 (q^2, m_{\pi\pi}^2, \theta_{\pi^{+}})&=&
 2 \beta_l  \sum_{J=0, ...} \sum_{J'=1, ..} \left[  Y_J^0 (\theta_{\pi^{+}}, 0)Y_{J'}^{-1}(\theta_{\pi^{+}}, 0) | A^J_{0}A^{J'*}_{\perp} | \sin(\delta_{0}^J -\delta_{\perp}^{J'})  \right],
  \end{eqnarray}
 \begin{eqnarray}
 I_9(q^2, m_{\pi\pi}^2, \theta_{\pi^{+}})
 &=& 2\beta_l \sum_{J=1, ...} \sum_{J'=1, ..} \left[  Y_J^{-1}(\theta_{\pi^{+}}, 0) Y_{J'}^{-1}(\theta_{\pi^{+}}, 0) | A^J_{\perp}A^{J'*}_{||} | \sin(\delta_{\perp}^J -\delta_{||}^{J'})  \right].
 \label{eq:angularCoefficients}
\end{eqnarray}

\end{appendix}



\begin{thebibliography}{11}
\bibitem{CLEO:2011ab}
  S.~Dobbs {\it et al.} [CLEO Collaboration],
  Phys.\ Rev.\ Lett.\  {\bf 110}, no. 13, 131802 (2013)
  [arXiv:1112.2884 [hep-ex]].


\bibitem{Wang:2009azc}
  W.~Wang and C.~D.~L\"{u},
  Phys.\ Rev.\ D {\bf 82}, 034016 (2010)
  [arXiv:0910.0613 [hep-ph]].


\bibitem{Achasov:2012kk}
  N.~N.~Achasov and A.~V.~Kiselev,
  Phys.\ Rev.\ D {\bf 86}, 114010 (2012)
  [arXiv:1206.5500 [hep-ph]].


\bibitem{Ablikim:2015orh}
  M.~Ablikim {\it et al.} [BESIII Collaboration],
  Phys.\ Lett.\ B {\bf 753}, 629 (2016)
  [arXiv:1507.08188 [hep-ex]].


\bibitem{Lu:2011jm}
  C.~D.~L\"u and W.~Wang,
  Phys.\ Rev.\ D {\bf 85}, 034014 (2012)
  [arXiv:1111.1513 [hep-ph]].


\bibitem{Meissner:2013pba}
  U.~G.~Mei{\ss}ner and W.~Wang,
  JHEP {\bf 1401}, 107 (2014)
  [arXiv:1311.5420 [hep-ph]].


\bibitem{Meissner:2013hya}
  U.~G.~Mei{\ss}ner and W.~Wang,
  Phys.\ Lett.\ B {\bf 730}, 336 (2014)
  [arXiv:1312.3087 [hep-ph]].


\bibitem{Wang:2015uea}
  W.~F.~Wang, H.~n.~Li, W.~Wang and C.~D.~L\"u,
  Phys.\ Rev.\ D {\bf 91}, no. 9, 094024 (2015)
  [arXiv:1502.05483 [hep-ph]].


\bibitem{Wang:2015paa}
  W.~Wang and R.~L.~Zhu,
  Phys.\ Lett.\ B {\bf 743}, 467 (2015)
  [arXiv:1502.05104 [hep-ph]].


\bibitem{Shi:2015kha}
  Y.~J.~Shi and W.~Wang,
  Phys.\ Rev.\ D {\bf 92}, no. 7, 074038 (2015)
  [arXiv:1507.07692 [hep-ph]].


\bibitem{Xie:2014tma}
  J.~J.~Xie, L.~R.~Dai and E.~Oset,
  Phys.\ Lett.\ B {\bf 742}, 363 (2015)
  [arXiv:1409.0401 [hep-ph]].


\bibitem{Sekihara:2015iha}
  T.~Sekihara and E.~Oset,
  Phys.\ Rev.\ D {\bf 92}, no. 5, 054038 (2015)
  [arXiv:1507.02026 [hep-ph]].


\bibitem{Oset:2016lyh}
  E.~Oset {\it et al.},
  Int.\ J.\ Mod.\ Phys.\ E {\bf 25}, 1630001 (2016)
  [arXiv:1601.03972 [hep-ph]].


\bibitem{Wang:2016rlo}
  W.~F.~Wang and H.~n.~Li,
  Phys.\ Lett.\ B {\bf 763}, 29 (2016)
  [arXiv:1609.04614 [hep-ph]].


\bibitem{Kang:2013jaa}
  X.~W.~Kang, B.~Kubis, C.~Hanhart and U.~G.~Mei{\ss}ner,
  Phys.\ Rev.\ D {\bf 89}, 053015 (2014)
  [arXiv:1312.1193 [hep-ph]].


\bibitem{Faller:2013dwa}
  S.~Faller, T.~Feldmann, A.~Khodjamirian, T.~Mannel and D.~van Dyk,
  Phys.\ Rev.\ D {\bf 89}, no. 1, 014015 (2014)
  [arXiv:1310.6660 [hep-ph]].


\bibitem{Niecknig:2015ija}
  F.~Niecknig and B.~Kubis,
  JHEP {\bf 1510}, 142 (2015)
  [arXiv:1509.03188 [hep-ph]].


\bibitem{Daub:2015xja}
  J.~T.~Daub, C.~Hanhart and B.~Kubis,
  JHEP {\bf 1602}, 009 (2016)
  [arXiv:1508.06841 [hep-ph]].


\bibitem{Albaladejo:2016mad}
  M.~Albaladejo, J.~T.~Daub, C.~Hanhart, B.~Kubis and B.~Moussallam,
  JHEP {\bf 1704}, 010 (2017)
  [arXiv:1611.03502 [hep-ph]].


\bibitem{Wirbel:1985ji}
  M.~Wirbel, B.~Stech and M.~Bauer,
  Z.\ Phys.\ C {\bf 29}, 637 (1985).


\bibitem{Scora:1995ty}
  D.~Scora and N.~Isgur,
  Phys.\ Rev.\ D {\bf 52}, 2783 (1995)
  [hep-ph/9503486].

\bibitem{Fajfer:2005ug}
  S.~Fajfer and J.~F.~Kamenik,
  Phys.\ Rev.\ D {\bf 72}, 034029 (2005)
  [hep-ph/0506051].

\bibitem{Verma:2011yw}
  R.~C.~Verma,
  J.\ Phys.\ G {\bf 39}, 025005 (2012)
  [arXiv:1103.2973 [hep-ph]].

\bibitem{Cheng:2003sm}
  H.~Y.~Cheng, C.~K.~Chua and C.~W.~Hwang,
  Phys.\ Rev.\ D {\bf 69}, 074025 (2004)
  [hep-ph/0310359].


\bibitem{Wu:2006rd}
  Y.~L.~Wu, M.~Zhong and Y.~B.~Zuo,
  Int.\ J.\ Mod.\ Phys.\ A {\bf 21}, 6125 (2006)
  [hep-ph/0604007].


\bibitem{Colangelo:2010bg}
  P.~Colangelo, F.~De Fazio and W.~Wang,
  Phys.\ Rev.\ D {\bf 81}, 074001 (2010)
  [arXiv:1002.2880 [hep-ph]].


\bibitem{Olive:2016xmw}
  C.~Patrignani {\it et al.} [Particle Data Group],
  Chin.\ Phys.\ C {\bf 40}, no. 10, 100001 (2016).


\bibitem{DeFazio:2001uc}
  F.~De Fazio and M.~R.~Pennington,
  Phys.\ Lett.\ B {\bf 521}, 15 (2001)
  [hep-ph/0104289].


\bibitem{Doring:2013wka}
  M.~D\"oring, U.~G.~Mei{\ss}ner and W.~Wang,
  JHEP {\bf 1310}, 011 (2013)
  [arXiv:1307.0947 [hep-ph]].


\bibitem{Diehl:2003ny}
  M.~Diehl,
  Phys.\ Rept.\  {\bf 388}, 41 (2003)
  [hep-ph/0307382].

\bibitem{Cheng:2005nb}
  H.~Y.~Cheng, C.~K.~Chua and K.~C.~Yang,
  Phys.\ Rev.\ D {\bf 73}, 014017 (2006)
  [hep-ph/0508104].
\bibitem{Bar:2012ce}
  O.~B\"ar and M.~Golterman,
  Phys.\ Rev.\ D {\bf 87}, no. 1, 014505 (2013)
  [arXiv:1209.2258 [hep-lat]].


\bibitem{Diehl:2005rn}
  M.~Diehl, A.~Manashov and A.~Sch\"afer,
  Phys.\ Lett.\ B {\bf 622}, 69 (2005)
  [hep-ph/0505269].



\bibitem{Diehl:1998dk}
  M.~Diehl, T.~Gousset, B.~Pire and O.~Teryaev,
  Phys.\ Rev.\ Lett.\  {\bf 81}, 1782 (1998)
  [hep-ph/9805380].


\bibitem{Diehl:2000uv}
  M.~Diehl, T.~Gousset and B.~Pire,
  Phys.\ Rev.\ D {\bf 62}, 073014 (2000)
  [hep-ph/0003233].

\bibitem{Colangelo:2015kha}
  G.~Colangelo, E.~Passemar and P.~Stoffer,
  Eur.\ Phys.\ J.\ C {\bf 75}, 172 (2015)
  doi:10.1140/epjc/s10052-015-3357-1
  [arXiv:1501.05627 [hep-ph]].


\bibitem{Gasser:1983yg}
  J.~Gasser and H.~Leutwyler,
  Annals Phys.\  {\bf 158}, 142 (1984).


\bibitem{Gasser:1984gg}
  J.~Gasser and H.~Leutwyler,
  Nucl.\ Phys.\ B {\bf 250}, 465 (1985).


\bibitem{Gasser:1984ux}
  J.~Gasser and H.~Leutwyler,
  Nucl.\ Phys.\ B {\bf 250}, 517 (1985).


\bibitem{Meissner:2000bc}
  U.~G.~Mei{\ss}ner and J.~A.~Oller,
  Nucl.\ Phys.\ A {\bf 679}, 671 (2001)
  [hep-ph/0005253].


\bibitem{Bijnens:1998fm}
  J.~Bijnens, G.~Colangelo and P.~Talavera,
  JHEP {\bf 9805}, 014 (1998)
  [hep-ph/9805389].


\bibitem{Bijnens:2003xg}
  J.~Bijnens and P.~Dhonte,
  JHEP {\bf 0310}, 061 (2003)
  [hep-ph/0307044].


\bibitem{Oller:1998hw}
  J.~A.~Oller, E.~Oset and J.~R.~Pel\'aez,
  Phys.\ Rev.\ D {\bf 59}, 074001 (1999)
  Erratum: [Phys.\ Rev.\ D {\bf 60}, 099906 (1999)]
  Erratum: [Phys.\ Rev.\ D {\bf 75}, 099903 (2007)]
  [hep-ph/9804209].


\bibitem{Oller:1997ti}
  J.~A.~Oller and E.~Oset,
  Nucl.\ Phys.\ A {\bf 620}, 438 (1997)
  Erratum: [Nucl.\ Phys.\ A {\bf 652}, 407 (1999)]
  [hep-ph/9702314].


\bibitem{Lahde:2006wr}
  T.~A.~L\"ahde and U.~G.~Mei{\ss}ner,
  Phys.\ Rev.\ D {\bf 74}, 034021 (2006)
  [hep-ph/0606133].


\bibitem{Oller:2007xd}
  J.~A.~Oller and L.~Roca,
  Phys.\ Lett.\ B {\bf 651}, 139 (2007)
  [arXiv:0704.0039 [hep-ph]].


\bibitem{Cabibbo:1965zzb}
  N.~Cabibbo and A.~Maksymowicz,
  Phys.\ Rev.\  {\bf 137}, B438 (1965)
  Erratum: [Phys.\ Rev.\  {\bf 168}, 1926 (1968)].


\bibitem{Pais:1968zza}
  A.~Pais and S.~B.~Treiman,
  Phys.\ Rev.\  {\bf 168}, 1858 (1968).



\bibitem{delAmoSanchez:2010fd}
  P.~del Amo Sanchez {\it et al.} [BaBar Collaboration],
  Phys.\ Rev.\ D {\bf 83}, 072001 (2011)
  [arXiv:1012.1810 [hep-ex]].




\bibitem{Lee:1992ih}
  C.~L.~Y.~Lee, M.~Lu and M.~B.~Wise,
  Phys.\ Rev.\ D {\bf 46}, 5040 (1992).





\bibitem{Ablikim:2014cea}
  M.~Ablikim {\it et al.} [BESIII Collaboration],
  Phys.\ Rev.\ D {\bf 89}, no. 5, 052001 (2014)
  [arXiv:1401.3083 [hep-ex]].


\bibitem{Asner:2008nq}
  D.~M.~Asner {\it et al.},
  Int.\ J.\ Mod.\ Phys.\ A {\bf 24}, S1 (2009)
  [arXiv:0809.1869 [hep-ex]].


\bibitem{Bediaga:2012py}
  R.~Aaij {\it et al.} [LHCb Collaboration],
  Eur.\ Phys.\ J.\ C {\bf 73}, no. 4, 2373 (2013)
  [arXiv:1208.3355 [hep-ex]].


\bibitem{Aushev:2010bq}
  T.~Aushev {\it et al.},
  arXiv:1002.5012 [hep-ex].

 \end{thebibliography}
\end{document}